\documentclass[12pt]{amsart}
\usepackage{latexsym,amsfonts,amsmath,amssymb}

\newcommand{\C}{{\mathcal{C}}}
\newcommand{\E}{{\mathcal{E}}}
\newcommand{\Q}{{\mathcal{Q}}}
\newcommand{\PP}{{\mathcal{P}}}
\newcommand{\LL}{{\mathcal{L}}}
\newcommand{\M}{{\mathcal{M}}}

\newcommand{\spec}{{\rm{spec}}}
\newcommand{\Tr}{{\rm{Tr}}}

\newcounter{smalllist}

\newtheorem{theorem}{Theorem}

\newtheorem{lemma}{Lemma}[section]
\newtheorem{prop}[lemma]{Proposition}
\newtheorem{coro}[lemma]{Corollary}
\theoremstyle{definition}

\newtheorem{remark}[lemma]{Remark}

\let\Re=\undefined\DeclareMathOperator{\Re}{Re}
\let\Im=\undefined\DeclareMathOperator{\Im}{Im}
\DeclareMathOperator{\diag}{diag}

\let\llldots=\ldots
\def\ldots{\llldots{}}

\numberwithin{equation}{section}

\begin{document}

\title[Lax pairs for the Ablowitz-Ladik system]{Lax pairs for the Ablowitz-Ladik system via orthogonal polynomials on the unit circle}
\author[I.~Nenciu]{Irina Nenciu}
\address{Irina Nenciu\\
         Mathematics 253-37\\
         Caltech\\
         Pasadena, CA 91125}
\email{nenciu@caltech.edu}

\begin{abstract}
In \cite{Simon2} Nenciu and Simon found that the analogue of the
Toda system in the context of orthogonal polynomials on the unit
circle is the defocusing Ablowitz-Ladik system. In this paper we use
the CMV and extended CMV matrices defined in \cite{CMV} and
\cite{Simon1,Simon2}, respectively, to construct Lax pair
representations for this system.
\end{abstract}

\maketitle

\section{Introduction}

The aim of this paper is to present new results concerning the
Ablowitz-Ladik (AL) system. More precisely, we use the connection
between the AL system and the theory of orthogonal polynomials on
the unit circle to construct Lax pairs associated to the
Hamiltonians in the defocusing AL hierarchy. Our main investigation
focuses on the periodic case, but these results translate to
corresponding statements in the finite and infinite cases.

We briefly introduce the main players. The defocusing AL equation
was defined in 1975--76 by Ablowitz and Ladik \cite{AL1,AL2} as a
space-discretization of the cubic nonlinear Schr\"odinger equation.
It reads:
\begin{equation}\label{ALE1}
-i\dot\alpha_n=\rho_n^2 (\alpha_{n+1}+\alpha_{n-1})-2\alpha_n,
\end{equation}
where $\alpha=\{\alpha_n\}\subset\mathbb{D}$ is a sequence of
complex numbers inside the unit disk and
$$
\rho_n^2=1-|\alpha_n|^2.
$$
The analogy with the continuous NLS becomes transparent if we
rewrite \eqref{ALE1} as
$$
-i\dot\alpha_n=\alpha_{n+1}-2\alpha_n+\alpha_{n-1}
-|\alpha_n|^2(\alpha_{n+1}+\alpha_{n-1}).
$$
Here, and throughout this paper, $\dot f$ will denote the time
derivative of the function $f$.

We note that the name ``Ablowitz-Ladik equation" that we use here
for \eqref{ALE1} is sometimes used for a more general equation that
was introduced in the same paper \cite{AL1}. Moreover, \eqref{ALE1}
also appears in the literature under the name IDNLS (integrable
discrete nonlinear Schr\"odinger equation). So far, the study of
this equation focused mainly around the inverse scattering
transform; see, for example, \cite[Chapter 3]{APT} and the
references therein. Other aspects of the Ablowitz-Ladik equations
have been further studied, for example, in \cite{GGH}, \cite{GH},
\cite{MEKL}, \cite{MSCE}, \cite{R}, \cite{V1}, and \cite{V2}.

We will try to understand \eqref{ALE1} from a different perspective:
that of the theory of orthogonal polynomials on the unit circle. We
concentrate on the periodic problem as it was the first one solved,
and our main results for the finite and infinite defocusing
Ablowitz-Ladik systems follow from the corresponding result in the
periodic case.

While more details can be found in Appendix~\ref{OPUC} and we shall
freely use all the notation introduced there, we present here some
of the main notions and relevant results. Let $\mu$ be a probability
measure on the unit circle. By applying the Gram-Schmidt procedure
to $1,z,z^2,\ldots$, one can define the monic orthogonal polynomials
$\{\Phi_n\}_{n\geq0}$. They obey a recurrence relation
$$
\Phi_{n+1}(z)=z\Phi_n(z)-\bar\alpha_n\Phi_n^*(z)
$$
for all $n\geq0$, where
$$
\Phi_n^*(z)=z^n\overline{\Phi_n(\tfrac{1}{\bar{z}})}
$$
is the reversed polynomial, and $\alpha=\{\alpha_n\}_{n\geq0}$ is
the sequence of Verblunsky coefficients. (The use of the same
notation as in \eqref{ALE1} is not a coincidence, as we will see.)
If we represent the operator of multiplication by $z$ on $L^2(d\mu)$
in an appropriate basis, we obtain a 5-diagonal unitary matrix
$$
\mathcal{C}=
\left(%
\begin{array}{cccccc}
  \bar\alpha_0 & \rho_0\bar\alpha_1    & \rho_0\rho_1    & 0 & 0 & \ldots \\
  \rho_0       & -\alpha_0\bar\alpha_1 & -\alpha_0\rho_1 & 0 & 0 & \ldots \\
  0 & \rho_1\bar\alpha_2 & -\alpha_1\bar\alpha_2 & \rho_2\bar\alpha_3 & \rho_2\rho_3 & \ldots \\
  0 & \rho_1\rho_2 & -\alpha_1\rho_2 & -\alpha_2\bar\alpha_3 & -\alpha_2\rho_3 & \ldots \\
  0 & 0 & 0 & \rho_3\bar\alpha_4 & -\alpha_3\bar\alpha_4 & \ldots \\
  \ldots & \ldots & \ldots & \ldots & \ldots & \ldots \\
\end{array}%
\right).
$$
This matrix was first discovered by Cantero, Moral, and Vel\'azquez
\cite{CMV}, and is called the CMV matrix. Furthermore, if the
Verblunsky coefficients are periodic, one can very naturally define
the extended CMV matrix $\mathcal E$ (as in \eqref{DefE}) and its
Floquet restrictions $\E_{(\beta)}$ to the periodic subspaces (see
Chapter~11 of \cite{Simon2}).

The theory of periodic Verblunsky coefficients was first studied by
Geronimus, and, more recently, by Peherstorfer and collaborators,
and Golinskii and collaborators (for detailed references to their
work, see \cite{Simon2}). Simon used the analogy with Hill's
equation to fully develop the theory for periodic Verblunsky
coefficients in \cite[Chapter~11]{Simon2}. In particular, he defines
the discriminant $\Delta(z)$ naturally associated to this periodic
problem and finds that

\begin{prop}[Simon]
Let $p$ {\rm(}the period of the coefficients{\rm)} be even. Let
$\{\alpha_j\}_{j=0}^{p-1}$ and $\{\gamma_j\}_{j=0}^{p-1}$ be two
elements of $\mathbb{D}^p$. The following are equivalent:
\begin{enumerate}
    \item $\Delta(z;\{\alpha_j\})=\Delta(z;\{\gamma_j\})$.
    \item $\prod_j (1-|\alpha_j|^2)=\prod_j (1-|\gamma_j|^2)$, and the eigenvalues of $\mathcal{E}_{(\beta)}(\{\alpha_j\}_{j=0}^{p-1})$ and $\mathcal{E}_{(\beta)}(\{\gamma_j\}_{j=0}^{p-1})$ coincide for one
    $\beta\in\partial\mathbb{D}$.
    \item The eigenvalues of $\mathcal{E}_{(\beta)}(\{\alpha_j\}_{j=0}^{p-1})$ and $\mathcal{E}_{(\beta)}(\{\gamma_j\}_{j=0}^{p-1})$ are equal for all
    $\beta\in\partial\mathbb{D}$.
    \item $\spec(\mathcal{E}(\{\alpha_j\}_{j=0}^{p-1}))=\spec(\mathcal{E}(\{\gamma_j\}_{j=0}^{p-1}))$.
\end{enumerate}
\end{prop}
When these conditions hold, we say that $\{\alpha_j\}_{j=0}^{p-1}$
and $\{\gamma_j\}_{j=0}^{p-1}$ are isospectral.

Next, we present two examples which represented a first step in
establishing the connection between OPUC and the AL system. For the
full computations which justify our claims, see Examples~11.1.4 and
5 in \cite{Simon2}.

\bigskip

\noindent {\bf Example 1} (Geronimus).  Let $\alpha\in\mathbb{D}$
and define $\alpha_j\equiv\alpha$ for all $j\geq0$.

The isospectral manifold in this case is a circle
$$
\{\alpha=(1-\rho^2)^{1/2}e^{i\theta}:\theta\in[0,2\pi]\}
$$
if $|\alpha|\neq0$, and a point (or a zero-dimensional torus),
$\alpha=0$, if $|\alpha|=0$.

\bigskip

\noindent {\bf Example 2} (Akhiezer). Consider $\alpha_{2j}=\alpha$
and $\alpha_{2j+1}=\alpha^{\prime}$, with
$\alpha,\alpha^{\prime}\in\mathbb{D}$ and $j\geq0$, to be periodic
Verblunsky coefficients with period $p=2$. Again the discriminant is
easily computable and
\begin{equation}\label{D2}
\Delta(e^{i\theta})=\frac{2}{\rho\rho'}[\cos(\theta)+\Re(\bar\alpha
\alpha')].
\end{equation}
Let $\theta_{\pm}\in[0,\pi)$ solve
$\cos(\theta_{\pm})=-\Re(\bar\alpha \alpha')\pm\rho\rho'$. Note that
$|\Re(\bar\alpha \alpha')|+\rho\rho'\leq1$, and hence there are
always solutions, with $0\leq\theta_+<\theta_-\leq\pi$. Hence
$|\Delta(e^{i\theta})|\leq2$ if and only if
$\pm\theta\in[\theta_+,\theta_-]$. We are interested in finding the
set of pairs $(\alpha,\alpha')\in\mathbb{D}^2$ which lead to a given
$\Delta$ of the form \eqref{D2}. This can be done explicitly, and
the conclusion is that
\begin{itemize}
    \item There are no open gaps for $|\alpha|=|\alpha'|=0$, and so the isospectral manifold is a point (0-dimensional
    torus).
    \item There is exactly one open gap when $\alpha=\pm\alpha'\neq0$, which leads to the isospectral manifold being a
    circle.
    \item There are two open gaps if and only if the isospectral manifold is a two-dimensional torus.
\end{itemize}
The two examples above suggest that $\mathbb{D}^p$ fibers into tori,
generically of real dimension $p$, half of the real dimension of
$\mathbb{D}^p$. This was proved by Simon in \cite{Simon2}.

Recall that a Hamiltonian vector field $V_H$ is called {\it
completely integrable} on a domain $D$ contained in a manifold of
real dimension $2n$ if there exist $n$ integrals of motion $H_1=H,
H_2,\ldots,H_n$ whose gradients are linearly independent on $D$ and
which Poisson commute. In this case, the Liouville-Arnold-Jost
Theorem (see \cite{AKN} and  \cite{Deift}) says that if $N=\bigcap_k
H_k^{-1}(c_k)$ is compact and connected, then it is an
$n$-dimensional torus. Finding the symplectic structure and the
integrable system naturally associated with periodic Verblunsky
coefficients and the notion of isospectrality defined above was the
main purpose of the work of Nenciu and Simon \cite[Ch.11, Section
11]{Simon2}, that we shall briefly describe here.

We begin by defining the symplectic structure. We are considering
the problem of periodic Verblunsky coefficients with period $p$. So
we are interested in a symplectic form on $\mathbb{D}^p$, which has
real dimension $2p$. Let
$\underline{\alpha}=(\alpha_0,\ldots,\alpha_{p-1})\in\mathbb{D}^p$,
and let $u_j=\Re \alpha_j$ and $v_j=\Im \alpha_j$ for all $0\leq
j\leq p-1$. Then we define our symplectic form by
\begin{equation}\label{DefSympl}
\omega=\frac12 \sum_{j=0}^{p-1} \frac{1}{\rho_j^2}\, du_j \wedge
dv_j.
\end{equation}
As all of the subsequent computations will involve only the
corresponding Poisson bracket, let us note that, for $f$ and $g$
functions on $\mathbb{D}^p$, we have
\begin{align}\label{DefPB}
\{f,g\} &=\frac12 \sum_{j=0}^{p-1} \rho_j^2 \left[\frac{\partial
f}{\partial u_j}\frac{\partial g}{\partial v_j}-
 \frac{\partial f}{\partial v_j}\frac{\partial g}{\partial u_j}\right]\\
&=i\sum_{j=0}^{p-1} \rho_j^2 \left[\frac{\partial f}{\partial
\bar\alpha_j}\frac{\partial g}{\partial \alpha_j}-
 \frac{\partial f}{\partial \alpha_j}\frac{\partial g}{\partial
 \bar\alpha_j}\right],
\end{align}
where for $z=u+iv\in \mathbb{D}$ we use the standard notation
$$
\frac{\partial}{\partial z}=\frac12 \left(\frac{\partial}{\partial
u}-i\frac{\partial}{\partial v}\right)\quad \text{and}\quad
\frac{\partial}{\partial \bar z}=\frac12
\left(\frac{\partial}{\partial u}+i\frac{\partial}{\partial
v}\right).
$$

\begin{lemma} The 2-form defined by \eqref{DefSympl} is a
symplectic form. Equivalently, the bracket \eqref{DefPB} obeys the
Jacobi identity and is nondegenerate.
\end{lemma}

\begin{proof}
$\omega$ is a sum of 2-forms, each of which acts only on one of the
variables $\alpha_j$ for $0\leq j\leq p-1$. But any 2-form is closed
in $\mathbb{R}^2$, and hence $\omega$ is closed. It is also
nondegenerate, since the function $\rho_j^{-2}$ is positive on
$\mathbb{D}^p$ for each $j$.
\end{proof}

The first result is
\begin{theorem}[Nenciu - Simon]\label{comm}
With the above Poisson bracket we have
\begin{equation}
\{\Delta(z),\Delta(w)\}=0
\end{equation}
for any $z,w\in\mathbb{C}$.
\end{theorem}
In particular, one has
\begin{coro}
The Hamiltonian flows generated by $\Delta(z)$ for
$z\in\partial\mathbb{D}$ and by $\prod_{j=0}^{p-1}\rho_j$ all
commute with each other and leave $\Delta(w)$ invariant.
\end{coro}

Theorem~\ref{comm} is proved using the expression of $\Delta$ in
terms of Wall polynomials and the recurrence relations that they
obey. For details, see Section~ 11.11 of \cite{Simon2} and the
references therein.

Moreover, the coefficients of the monic polynomial
$$
z^{p/2}\Big(\prod_{j=0}^{p-1}\rho_j\Big) \Delta(z)
$$
come in complex conjugate pairs: $c_j=\bar c_{p-j}$. If we also take
into account the fact that $c_0=c_p=1$ and $c_{p/2}$ is real, we get
that the real and imaginary parts of the $c_j$ with $1\leq j\leq
\frac{p}{2}-1$, together with $c_{p/2}$ and
$\prod_{j=0}^{p-1}\rho_j^2$, form a set of $p$ commuting integrals.
Note that the real dimension of the space $\mathbb{D}^p$ of the
Verblunsky coefficients is $2p$, twice the number of the
Hamiltonians described above.

The next natural question concerns finding Lax pairs associated to
these commuting Hamiltonians.

\begin{remark}\label{LPgeneric}
Recall that finding a Lax pair representation
\begin{equation}\label{genericLP}
\dot L=[L,P]
\end{equation}
for an evolution equation allows one to identify the eigenvalues of
$L$ as conserved quantities. This is usually done in the case when
$L$ is selfadjoint, but essentially the same proof works in the case
that we are interested in, when $L$ is unitary.

Indeed, let $\lambda\in S^1$ be an eigenvalue of $L$, a unitary
matrix, and $\phi$ the corresponding unit eigenvector. Then
$$
\lambda=(L\phi,\phi)
$$
and hence
$$
\dot{\lambda}=(\dot{L}\phi,\phi)+(L\dot{\phi},\phi)+(L\phi,\dot{\phi}).
$$
Note that
\begin{align*}
(L\dot{\phi},\phi)+(L\phi,\dot{\phi})&=(\dot{\phi},L^*\phi)+(L\phi,\dot{\phi})\\
&=(\dot{\phi},\bar\lambda\phi)+(\lambda\phi,\dot{\phi})\\
&=\lambda[(\dot{\phi},\phi)+(\phi,\dot{\phi})]\\
&=\lambda\dot{(\phi,\phi)}\\
&=0,
\end{align*}
as $(\phi,\phi)\equiv1$. So,
\begin{align*}
\dot{\lambda}&=(\dot{L}\phi,\phi)=([L,P]\phi,\phi)\\
&=(P\phi,L^*\phi)-(PL\phi,\phi)\\
&=(P\phi,\bar\lambda\phi)-(\lambda P\phi,\phi)=0.
\end{align*}
\end{remark}

As it turns out, the coefficients $c_j$ for which Nenciu and Simon
obtained Poisson commutativity are not the Hamiltonians we will be
working with here, while being closely related to them.

A consequence of Theorem 11.2.2 and formula (11.2.17) of
\cite{Simon2} is
\begin{equation}\label{DeltaK}
\det(z-\Q_{(1)})=z^{p/2}\Big(\prod_{j=0}^{p-1}\rho_j\Big)[\Delta(z)-2],
\end{equation}
where $\Q_{(1)}$ is the restriction of $\E$ to the space of
$p$-periodic $l^{\infty}$ sequences (see the definition of
$\Q_{(d)}$ in Section~\ref{MainResultsP}). In particular, this shows
that $\prod_{j=0}^{p-1}\rho_j$ and the coefficients of $\Delta$
Poisson commute if and only if $\prod_{j=0}^{p-1}\rho_j$ and the
real and imaginary parts of the traces of the first $\frac{p}{2}$
powers of $\Q_{(1)}$ Poisson commute. These are (essentially, see
Proposition~\ref{Kof1}) the Hamiltonians we will be working with.
They are the natural functions to consider by the fact proved above,
that existence of Lax pairs implies conservation of the eigenvalues,
and hence of the traces of powers of the Lax matrix.

The organization of the paper is as follows: In
Section~\ref{MainResultsP} we define the objects involved in the
periodic problem and present the main result, Theorem~\ref{LP}, and
its consequences. Section~\ref{ProofP} contains the ideas of the
proof of the main theorem, Theorem~\ref{LP}, while in
Appendix~\ref{Comp} we give the full computations involved in this
proof. Sections~\ref{Finite} and \ref{Infinite} deal with the finite
and infinite cases, respectively. Finally, Appendix~\ref{OPUC} gives
the necessary background on the theory of orthogonal polynomials on
the unit circle.

\section{The Main Results in the Periodic Case}\label{MainResultsP}

We must first define our Hamiltonians $K_n$. Essentially, they are
traces per volume of the powers of the extended CMV matrix $\E$.

Consider the periodic Ablowitz-Ladik problem with period $p$. If $p$
is even, then let $\E$ be the extended CMV matrix associated to
these $\alpha$'s; if $p$ is odd, think of the sequence of Verblunsky
coefficients as having period $2p$ and thus define the extended CMV
matrix $\E$.

For each $n\geq1$, we define the Hamiltonians we will be working
with as:
\begin{equation}\label{Kndefn}
K_n=\frac{1}{n}\sum_{k=0}^{p-1} \E^n_{kk}\,.
\end{equation}
For $n=0$, we set
$$
K_0=\prod_{j=0}^{p-1}\rho_j^2.
$$

Finally, for $\mathcal A$ a doubly-infinite matrix, we set
$\mathcal{A}_+$ as the matrix with entries
$$
(\mathcal{A}_+)_{jk}=\left\{%
\begin{array}{ll}
    \mathcal{A}_{jk}, & \quad\hbox{if}\,\,j<k;\\
    \tfrac12\mathcal{A}_{jj}, & \quad\hbox{if}\,\,j=k;\\
    0, & \quad\hbox{if}\,\,j>k.\\
\end{array}%
\right.
$$

The central theorem of our paper is:
\begin{theorem}\label{LP}
The Lax pairs for the $n^{\rm{th}}$ Hamiltonian of the periodic
defocusing Ablowitz-Ladik system are given by
\begin{equation}\label{LPK}
\{\E,K_{n}\}=[\E,i\E^{n}_+]
\end{equation}
and
\begin{equation}\label{LPKbar}
\{\E,\bar{K}_{n}\}=[\E,i(\E^{n}_+)^*]
\end{equation}
for all $n\geq1$.
\end{theorem}

Here we use $\{\E,f\}$ to denote the doubly-infinite matrix with
$(j,k)$ entry $\{\E_{jk},f\}$; also, $\E^n_+$ denotes $(\E^n)_+\,$.

\begin{remark} The form of Theorem~\ref{LP}, and the main idea of the proof
were inspired by the analogous result of van Moerbeke~\cite{vanM}
for the periodic Toda lattice. But none of the two results implies
the other.

Moreover, in the case of the Toda lattice, the necessary
calculations are very simple due to the tri-diagonal, symmetric
nature of the Jacobi matrices naturally associated to that problem.
The analogue on the circle are CMV matrices, whose more complicated
structure makes proving this result computationally much more
involved.
\end{remark}

But we are dealing with a finite dimensional problem, so we are
interested in finding appropriate finite dimensional spaces to which
we can restrict the operators in \eqref{LPK} and \eqref{LPKbar}.
Also, we want to express the Hamiltonians $K_n$ in terms of these
restrictions.

The following lemma is an immediate consequence of the structure of
$\E$; it can easily be proven by induction whenever $\E$ can be
defined.

\begin{lemma}\label{nzE}
Let $n\geq1$ be an integer. Then $\E_{j,k}^n$ is identically zero as
a function of the Verblunsky coefficients  if one of the following
holds:
\begin{itemize}
    \item[] $|j-k|\geq2n+1$
    \item[or] $j-k=2n$ and $j$ and $k$ are even
    \item[or] $j-k=-2n$ and $j$ and $k$ are odd.
\end{itemize}
In particular, the number of entries which are not identically zero
{\rm(}as functions of the $\alpha$'s{\rm)} on any row of $\mathcal
{E}^n$ is bounded by $4n$.
\end{lemma}

Recall that the definition of $\E$ depends on the parity of the
period $p$. This explains why we need to study the cases $p$ even
and $p$ odd separately.

\bigskip

Let us first consider the case of the period $p$ being even. We
denote by $X_{(d)}$ the subspace of $l^{\infty}(\mathbb{Z})$
$$
X_{(d)}=\{u\in l^{\infty}(\mathbb{Z})\,|\, u_{m+dp}=u_m\}
$$
of sequences of period $dp$. As the Verblunsky coefficients are
periodic with period $p$, we find that $\E_{j+p,k+p}=\E_{j,k}$ for
any $j,k\in \mathbb{Z}$, and hence $\E^n$ restricts to $X_{(d)}$ for
all $n\in\mathbb{Z}$ and $d\geq1$. Moreover, if we denote by
$\xi_k^{(d)}$, $k=0,\ldots,dp-1$, the $l^{\infty}(\mathbb{Z})$
vector given by
$$
(\xi_k^{(d)})_j=1\,\,\text{when}\,\,j\equiv
k\,(\text{mod}\,dp),\,\,\text{and}\,\,0\,\,\text{otherwise,}
$$
we have that $\{\xi_0^{(d)}, \xi_1^{(d)},\ldots,\xi_{dp-1}^{(d)}\}$
is a basis in $X_{(d)}$, and
$$
(\E^n\xi_k^{(d)})_{j+p}=(\E^n\xi_k^{(d)})_{j}=\sum_{l\in\mathbb{Z}}\E^n_{j,k+ldp}\,.
$$
Notice that this sum has only a finite number of nonzero terms for
any choice of $n,j,k,p$, and $d$.

Let us denote by $\Q_{(d)}$ the matrix representation of the
restriction $\E\upharpoonright X_{(d)}$ in the basis $\{\xi_0^{(d)},
\xi_1^{(d)},\ldots,\xi_{dp-1}^{(d)}\}$. Then the matrix representing
$\E^n\upharpoonright X_{(d)}$ in the same basis is $\Q^n_{(d)}$,
whose entries are given by
\begin{equation}\label{Qdn}
\Q^n_{(d),jk}=\sum_{l\in\mathbb{Z}}\E^n_{j,k+ldp}
\end{equation}
for $0\leq j,k\leq dp-1$.

\begin{lemma}
For $dp\geq 2n+1$, we have that
$$
\frac{1}{d}{\rm Tr}(\Q_{(d)}^n)
$$
is independent of $d$ and equals $K_n\,$.
\end{lemma}

\begin{proof}
$$
\frac{1}{d}{\rm Tr}(\Q_{(d)}^n)=\frac{1}{d}\sum_{k=0}^{dp-1}
\Q^n_{(d),kk}\,.
$$
From Lemma~\ref{nzE} we know that $\E^n_{jk}=0$ for $|j-k|>2n$. So
for $dp\geq 2n+1$ we get
$$
\Q^n_{(d),kk}=\sum_{l\in\mathbb{Z}}\E^n_{k,k+ldp}=\E^n_{kk}\,.
$$
From this and periodicity we can conclude that
$$
\frac{1}{d}{\rm Tr}(\Q_{(d)}^n)=\frac{1}{d}\sum_{k=0}^{dp-1}
\E^n_{kk}=\sum_{k=0}^{p-1} \E^n_{kk}=nK_n
$$
is indeed independent of $d$.
\end{proof}

\bigskip

If $p$ is odd, we consider the same objects as above, with the extra
constraint that $dp$, and hence $d$, must always be even. Recall
that in this case we define $\E$ by thinking of the Verblunsky
coefficients as having period $2p$. For $d$ even, we can then define
$X_{(d)}$ and $\Q_{(d)}$ as above, while always keeping in mind that
we can use the results we just proved for $dp=\frac{d}{2}\cdot2p$.

Therefore, if $d$ is even and large enough, we have that
$$
\frac{2}{d}{\rm Tr}(\Q_{(d)}^n)=\frac{2}{d}\sum_{k=0}^{dp-1}
\E^n_{kk}=\sum_{k=0}^{2p-1} \E^n_{kk}\,.
$$
The last observation we need to make is that in this case the
entries of $\E$ obey
$$
\E_{jk}=\E_{k+p,j+p}\,.
$$
This comes from the fact that
$$
\LL_{j+p,k+p}=\M_{jk}\,,
$$
and that $\LL$ and $\M$ are symmetric. Hence
$$
\E_{k+p,j+p}=\sum_{l\in\mathbb{Z}}\LL_{k+p,l}\M_{l,j+p}
=\sum_{l\in\mathbb{Z}} \LL_{kl}\M_{lj}
=\sum_{l\in\mathbb{Z}}\LL_{lk}\M_{jl}=\E_{jk}\,,
$$
as claimed. A straightforward induction shows that
$$
\E^n_{k+p,j+p}=\E^n_{jk}
$$
for all $n$, and hence
$$
\frac{1}{d}{\rm Tr}(\Q_{(d)}^n)=\frac12\sum_{k=0}^{2p-1} \E^n_{kk}
=\sum_{k=0}^{p-1} \E^n_{kk}=nK_n
$$
also holds for $p$ odd, as long as $dp$ is even and $dp\geq 2n+1$.

\bigskip

So we proved that, with $K_n$ defined as in \eqref{Kndefn}, we have
\begin{equation}\label{KnQd}
K_n=\frac{1}{dn}{\rm Tr}(\Q_{(d)}^n)
\end{equation}
for $dp$ even and greater than $2n+1$.

Let us note that relations \eqref{LPK} and \eqref{LPKbar} hold in
the sense of bounded operators on $l^{\infty}(\mathbb{Z})$.
Moreover, all the matrices in these relations obey the same
periodicity conditions as $\E$, so it makes sense to restrict
\eqref{LPK} and \eqref{LPKbar} to $X_{(d)}$ for $d\geq1$. By doing
this we get

\begin{coro}\label{LPd}
For all $d\geq1$, with $dp$ even, and $n\geq1$, we have
\begin{equation}
\{\Q_{(d)},K_{n}\}=[\Q_{(d)},i\Q_{(d),+}^{n}]
\end{equation}
and
\begin{equation}
\{\Q_{(d)},\bar K_{n}\}=[\Q_{(d)},i(\Q_{(d),+}^{n})^*],
\end{equation}
where we denote by $\Q_{(d),+}^{n}$ the matrix representation of
$(\E^n)_+\upharpoonright X_{(d)}$ in the basis $\{\xi_0^{(d)},
\xi_1^{(d)},\ldots,\xi_{dp-1}^{(d)}\}$.
\end{coro}

Note that $\Q_{(d),+}^{n}$ is not an upper triangular matrix, as it
contains entries which are generically nonzero in its lower left
corner.

Let us make an observation that will explain why we cannot simply
use the traces of powers of $\Q_{(1)}$ even if $p$ is even, but also
that we are not changing by much the Hamiltonians we are most
interested in:

\begin{prop}\label{Kof1}
For $p$ even and $1\leq n\leq \frac{p}{2}-1$, we have that
$$
K_n=\frac{1}{n}{\rm{Tr}}(\Q_{(1)}^n),
$$
but
$$
\frac{2}{p}{\rm{Tr}}(\Q_{(1)}^{p/2})=K_{p/2}+2K_0^{1/2}.
$$
\end{prop}

\begin{proof} From formula~\eqref{Qdn} and Lemma~\ref{nzE} we see that, for
$n\leq\frac{p}{2}-1$,
$$
\Q^n_{(1),jj}=\sum_{l\in\mathbb{Z}}\E^n_{j,j+lp}=\E^n_{jj}
$$
for all $j=0,\ldots,p-1$. This follows since, for $|l|\geq1$,
$$
|j-(j+lp)|\geq p \geq 2n+1.
$$
Hence, using \eqref{Kndefn},
$$
\frac{1}{n}\Tr(\Q_{(1)}^n)=\frac{1}{n}\sum_{j=0}^{p-1}
\E^n_{jj}=K_n.
$$

If $n=\frac{p}{2}$ and $j$ even, the formulae
\eqref{Qdn},\eqref{Eeven}, \eqref{Eodd}, and periodicity of the
Verblunsky coefficients imply that
\begin{align*}
\Q^n_{(1),jj}&=\sum_{l\in\mathbb{Z}}\E^n_{j,j+lp}\\
             &=\E^n_{jj}+\E^n_{j,j+p}\\
             &=\E^n_{jj}+\prod_{k=0}^{p-1}\rho_k
\end{align*}
and
\begin{align*}
\Q^n_{(1),j+1,j+1}&=\sum_{l\in\mathbb{Z}}\E^n_{j+1,j+1+lp}\\
                  &=\E^n_{j+1,j+1}+\E^n_{j+1,j+1-p}\\
                  &=\E^n_{j+1,j+1}+\prod_{k=0}^{p-1}\rho_k.
\end{align*}
Therefore,
$$
\frac{2}{p}{\rm{Tr}}(\Q_{(1)}^{p/2})=\frac{2}{p}\sum_{j=0}^{p-1}
\E^n_{jj}+\frac{2p}{p}\prod_{k=0}^{p-1}\rho_k=K_{p/2}+2K_0^{1/2},
$$
as claimed.
\end{proof}

\begin{remark}
An easy computation shows that
$$
\{\alpha_j,2\Re(K_1)\}=i\rho_j^2(\alpha_{j-1}+\alpha_{j+1})
$$
and
$$
\{\alpha_j,\log(K_0)\}=i\alpha_j
$$
for all $0\leq j\leq p-1$. Hence \eqref{ALE1}, the periodic
defocusing Ablowitz-Ladik equation, is the evolution of the
Verblunsky coefficients under the flow generated by the Hamiltonian
$2\Re(K_1)-2\log(K_0)$.
\end{remark}

From Theorem~\ref{LP} and Corollary~\ref{LPd} we can immediately
conclude that

\begin{coro}
The Lax pairs for the Hamiltonians $\Re(K_n)$ and $\Im(K_n)$,
$n\geq1$, are given by
\begin{equation}\label{LPR}
\{\E,2\Re(K_{n})\}=[\E,i\E^{n}_++i(\E^{n}_+)^*]
\end{equation}
and
\begin{equation}\label{LPI}
\{\E,2\Im(K_{n})\}=[\E,\E^{n}_+-(\E^{n}_+)^*],
\end{equation}
while the corresponding statements for $\Q_{(d)}$, $d\geq1$ and $dp$
even, are given by
\begin{equation}\label{LPRd}
\{\Q_{(d)},2\Re(K_{n})\}=[\Q_{(d)},i\Q_{(d),+}^{n}+i(\Q_{(d),+}^{n})^*]
\end{equation}
and
\begin{equation}\label{LPId}
\{\Q_{(d)},2\Im(K_{n})\}=[\Q_{(d)},\Q_{(d),+}^{n}-(\Q_{(d),+}^{n})^*]\,.
\end{equation}
\end{coro}

In particular, relations \eqref{LPRd} and \eqref{LPId}, together
with Remark~\ref{LPgeneric} and \eqref{KnQd}, imply that

\begin{coro}\label{CommKP}
$$
\{K_n,\Re(K_m)\}=\{K_n,\Im(K_m)\}=0,
$$
and hence
$$
\{K_n,K_m\}=\{K_n,\bar K_m\}=0.
$$
\end{coro}

Define the doubly-infinite matrix $\PP$ by
$$
\PP_{lm}=(-1)^l\delta_{lm}\frac{i}{2}\big(\prod_{k=0}^{p-1}\rho_k^2\big).
$$

\begin{prop}\label{LPK0}
The Lax pair representation for the flow generated by
$K_0=\prod_{j=0}^{p-1} \rho_j^2$ is given by
\begin{equation}\label{LPforK0}
\{\E,K_0\}=[\E,\PP].
\end{equation}
In particular, we can conclude that
\begin{equation}\label{commK0Kn}
\{K_0,K_n\}=\{K_0,\bar K_n\}=0,
\end{equation}
or, equivalently,
\begin{equation}\label{commK0RIKn}
\{K_0,2\Re(K_n)\}=\{K_0,2\Im(K_n)\}=0.
\end{equation}
\end{prop}

\begin{proof}
The Lax pair representation \eqref{LPforK0} is checked by a
straightforward computation. It is based on the fact that the flow
generated by $K_0$ rotates all the $\alpha$'s by the same angle
$$
\{\alpha_j,K_0\}=iK_0\alpha_j\,,
$$
while
$$
[\E,\PP]_{j,k}=\E_{j,k}(\PP_{k,k}-\PP_{j,j}).
$$

The Poisson commutation relations \eqref{commK0Kn} and
\eqref{commK0RIKn} follow, as in the previous cases, by restricting
the Lax pair to periodic subspaces and concluding that the flow
preserves eigenvalues, and hence traces.
\end{proof}

From \eqref{DeltaK}, Corollary~\ref{CommKP}, and
Proposition~\ref{LPK0}, we immediately get that
$\prod_{j=0}^{p-1}\rho_j$ and the coefficients $c_k$ of
$z^{p/2}\Big(\prod_{j=0}^{p-1}\rho_j\Big)[\Delta(z)-2]$ Poisson
commute. Note also that, by \eqref{DeltaK}, we see that the
connection between the $K$'s and the $c$'s cannot be explicitly
written down. Hence one cannot write simple Lax pairs in terms of
$\E$ for the flows generated by the $c$'s.


\section{The Periodic Case: Proof of
Theorem~\ref{LP}}\label{ProofP}

The main technical ingredient in the proof of Theorem~\ref{LP} is
the following:

\begin{lemma}\label{PartialK}
For all $n\geq0$ and $j$ even, we have

\begin{equation}\label{PK1}
\begin{aligned}
\frac{\partial K_{n+1}}{\partial \alpha_j}=
&-\frac{\bar\alpha_j\bar\alpha_{j+1}}{2\rho_j} \E^n_{j+1,j}
-\frac{\bar\alpha_j\rho_{j+1}}{2\rho_j}\E^n_{j+2,j}
-\frac{\bar\alpha_j\rho_{j-1}}{2\rho_j}\E^n_{j-1,j+1}\\
&+\frac{\bar\alpha_j\alpha_{j-1}}{2\rho_j}\E^n_{j,j+1}
-\bar\alpha_{j+1}\E^n_{j+1,j+1} -\rho_{j+1}\E^n_{j+2,j+1}
\end{aligned}
\end{equation}

\begin{equation}\label{PK2}
\begin{aligned}
\frac{\partial K_{n+1}}{\partial \bar\alpha_j}=
&\rho_{j-1}\E^n_{j-1,j} -\alpha_{j-1}\E^n_{j,j}
-\frac{\alpha_j\bar\alpha_{j+1}}{2\rho_j} \E^n_{j+1,j}\\
&-\frac{\alpha_j\rho_{j+1}}{2\rho_j}\E^n_{j+2,j}
-\frac{\alpha_j\rho_{j-1}}{2\rho_j}\E^n_{j-1,j+1}
+\frac{\alpha_j\alpha_{j-1}}{2\rho_j}\E^n_{j,j+1}
\end{aligned}
\end{equation}

\begin{equation}\label{PK3}
\begin{aligned}
\frac{\partial K_{n+1}}{\partial \alpha_{j-1}}=
&-\frac{\bar\alpha_{j-1}\rho_{j-2}}{2\rho_{j-1}} \E^n_{j,j-2}
+\frac{\bar\alpha_{j-1}\alpha_{j-2}}{2\rho_{j-1}}\E^n_{j,j-1}
-\frac{\bar\alpha_{j-1}\bar\alpha_{j}}{2\rho_{j-1}}\E^n_{j-1,j}\\
&-\frac{\bar\alpha_{j-1}\rho_{j}}{2\rho_{j-1}}\E^n_{j-1,j+1}
-\bar\alpha_{j}\E^n_{j,j} -\rho_{j}\E^n_{j,j+1}
\end{aligned}
\end{equation}

\begin{equation}\label{PK4}
\begin{aligned}
\frac{\partial K_{n+1}}{\partial \bar\alpha_{j-1}}=
&\rho_{j-2}\E^n_{j-1,j-2} -\alpha_{j-2}\E^n_{j-1,j-1}
-\frac{\alpha_{j-1}\rho_{j-2}}{2\rho_{j-1}} \E^n_{j,j-2}\\
&-\frac{\alpha_{j-1}\bar\alpha_{j}}{2\rho_{j-1}}\E^n_{j-1,j}
-\frac{\alpha_{j-1}\rho_{j}}{2\rho_{j-1}}\E^n_{j-1,j+1}
+\frac{\alpha_{j-1}\alpha_{j-2}}{2\rho_{j-1}}\E^n_{j,j-1}\,.
\end{aligned}
\end{equation}
\end{lemma}

\begin{remark}
Note that, for any $n\geq1$ and $0\leq j\leq p-1$, we have
$$
\frac{\partial \bar K_n}{\partial
\beta_j}=\overline{\left(\frac{\partial K_n}{\partial
\bar\beta_j}\right)}
$$
and hence one can easily find the derivatives of $\bar K_n$ with
respect to $\alpha_j$ and $\bar\alpha_j$ from Lemma~\ref{PartialK}
\end{remark}

\begin{proof}
The proof reduces to direct computations once one notices that, by
invariance of the trace under circular permutations,
\begin{align}\label{PKgen}
\frac{\partial K_{n+1}}{\partial \beta_j}&=\frac{1}{(n+1)d} {\rm
Tr}\left( \frac{\partial \Q_{(d)}}{\partial \beta_j}\Q_{(d)}^n
+\Q_{(d)} \frac{\partial \Q_{(d)}}{\partial \beta_j}\Q_{(d)}^{n-1}
+\cdots+\Q_{(d)}^n\frac{\partial \Q_{(d)}}{\partial
\beta_j}\right)\\
&=\frac{1}{d}{\rm Tr}\left(\frac{\partial \Q_{(d)}}{\partial
\beta_j}\Q_{(d)}^n\right).
\end{align}

We give here the complete proof of \eqref{PK1}; \eqref{PK2} through
\eqref{PK4} can be found in a similar way.

Notice that, for $j$ even, $\alpha_j$ appears in exactly $6d$
entries of $\Q_{(d)}$. So \eqref{PK1} follows by periodicity and by
a straightforward computation from \eqref{PKgen}:
\begin{equation*}
\begin{aligned}
\frac{\partial K_{n+1}}{\partial \alpha_j}
&=\frac{1}{d}\sum_{k,l}\frac{\partial\Q_{(d),kl}}{\partial\alpha_j}
\Q^n_{(d),lk}\\
&\begin{aligned}=&-\frac{\bar\alpha_j}{\rho_j}\bar\alpha_{j+1}\E^n_{j+1,j} -\frac{\bar\alpha_j}{\rho_j}\rho_{j+1}\E^n_{j+2,j} -\frac{\bar\alpha_j}{\rho_j}\rho_{j-1}\E^n_{j-1,j+1}\\
&+\frac{\bar\alpha_j}{\rho_j}\alpha_{j-1}\E^n_{j,j+1}-\bar\alpha_{j+1}\E^n_{j+1,j+1}-\rho_{j+1}\E^n_{j+2,j+1}.
\end{aligned}
\end{aligned}
\end{equation*}
\end{proof}

Before we embark on the proof of the main theorem, we provide
another preliminary result; while the statement is almost certainly
not new, we give a proof for the reader's convenience.

Consider an $N\times N$ matrix $A$ having the following
\emph{stair-shape}
$$
A=
\left(%
\begin{array}{ccccc}
  \star & 0      & 0      & \cdots   & 0 \\
  \star & \star  & 0      & \cdots   & 0 \\
  \vdots& \vdots & \vdots & \ddots   & \vdots\\
  \star & \star  & \star  & \cdots   & 0 \\
  \star & \star  & \star  & \cdots   & 0 \\
\end{array}%
\right),
$$
where the stars and 0's represent rectangular matrix blocks.
Formally, that means that for any row number $i$ there exists a
column number $j(i)$ so that $A_{ij}=0$ for all $j>j(i)$, and the
function $i\mapsto j(i)$ is non-decreasing. In particular, it is
also true that for any column $j$ there exists a row $i(j)$ so that
$A_{ij}=0$ for $i<i(j)$. Note in passing that $j(i)$ and $i(j)$ are
not equal.

We will say, somewhat informally, that another matrix $\tilde{A}$
\emph{has the same shape as $A$} if $\tilde{A}_{ij}=0$ whenever
$j>j(i)$ for all $i$.

\begin{lemma}\label{Shape}
Let $A$ be a matrix as above and $B$ an arbitrary $N\times N$
matrix. Then
\begin{equation}
[A,B_+]_{ij}=[A,B]_{ij}
\end{equation}
for all $(i,j)$ with $j>j(i)$. This implies that, for the same
indices $(i,j)$ with $j>j(i)$, we have
\begin{equation}
[A,B_-]_{ij}=0.
\end{equation}
\end{lemma}

\begin{remark} Note that:
\begin{itemize}
  \item If $A$ and $B$ commute, then the commutators
$[A,B_+]$ and $[A,B_-]$ have the same shape as $A$.
  \item Also, by
transposing these equations, we obtain the same type of result for
``lower triangle shapes."
  \item The same type of result holds for doubly-infinite matrices.
  In particular, if $\mathcal A$ and $\mathcal B$ are two doubly
  infinite, stair-shaped matrices such that the commutator $[\mathcal A,\mathcal
  B]$ makes sense and equals 0, then the commutators $[\mathcal A,\mathcal B_+]$ and $[\mathcal A,\mathcal B_-]$ are
  themselves stair-shaped.
\end{itemize}
\end{remark}

\begin{proof}
We proceed by direct computation: Let $(i,j)$ be an index so that
$j>j(i)$; equivalently, $i<i(j)$. Then
\begin{align*}
[A,B_+]_{ij} &=\sum_{k}A_{ik}B_{+,kj}-\sum_k B_{+,ik}A_{kj}\\
&=\sum_{k\leq j(i)<j}A_{ik}B_{+,kj}-\sum_{i<i(j)\leq k} B_{+,ik}A_{kj}\\
&=\sum_{k}A_{ik}B_{kj}-\sum_k B_{ik}A_{kj}\\
&=[A,B]_{ij}.
\end{align*}
Since $B_-=B-B_+$, we get that
$$
[A,B_-]=[A,B]-[A,B_+]
$$
and so the second relation is just a consequence of the first one.
\end{proof}

We are now ready to prove Theorem~\ref{LP}.

\begin{proof}
We will first deal with relation~\eqref{LPK} for $n+1$, $n\geq0$:
$$
\{\E,K_{n+1}\}=i[\E,\E^{n+1}_+]
$$
The left-hand side matrix has two types of entries: the ones outside
the shape of a CMV matrix, which are identically zero, and the ones
inside the shape.

The entries outside the shape are dealt with immediately by applying
Lemma~\ref{Shape}. Indeed, $\E$ and $\E^n$ are doubly-infinite
matrices, and they commute; hence, by the third observation above,
the commutator $[\E,\E^{n+1}_+]$ has the same shape as $\E$.

We are now left with the entries $(j,k)$ which are inside the shape.
Before we start computing, we make a short observation. Consider the
doubly-infinite matrix $\mathcal U$ given by
$$
\mathcal U_{jk}=\delta_{j,k+1}
$$
for all $j,k\in\mathbb{Z}$. In other
words, $\mathcal U$ is the left-shift on $l^{\infty}(\mathbb{Z})$ in
the usual basis. Note that for a doubly-infinite matrix $\mathcal B$
we have
$$
(\mathcal{U^*BU})_{jk}=B_{j-1,k-1}\quad\text{and}\quad
(\mathcal{UBU^*})_{jk}=B_{j+1,k+1}.
$$

Consider $\E=\E(\{\alpha_j\})$ to be a doubly-infinite CMV matrix.
We know that $\E=\mathcal{\tilde{L}\tilde{M}}$ with
\begin{equation*}
\tilde{\LL}=\diag\bigl(\ldots,\Theta_0,\Theta_2,\Theta_4,\ldots\bigr)
\end{equation*}
and
\begin{equation*}
\tilde{\M}=\diag\bigl(\ldots,\Theta_{-1},\Theta_1,\Theta_3,\ldots\bigr).
\end{equation*}
It is easily seen that
\begin{equation*}
\mathcal{U^*\tilde{L}}(\{\alpha_j\})^t\mathcal U=
\tilde{\M}(\{\alpha_{j-1}\}) \quad\text{and}\quad
\mathcal{U^*\tilde{M}}(\{\alpha_j\})^t\mathcal U=
\tilde{\LL}(\{\alpha_{j-1}\}),
\end{equation*}
which implies that
\begin{equation}
\mathcal{U^*E}(\{\alpha_j\})^t\mathcal U=\E(\{\alpha_{j-1}\})
\end{equation}
is also a doubly-infinite CMV matrix. The same is true for
$$
\mathcal{UE}(\{\alpha_j\})^t\mathcal U^*=\E(\{\alpha_{j+1}\}).
$$

We use the notation \eqref{LPK}$_{kl}$ for the $(k,l)$ entry of
relation \eqref{LPK}, and similarly for \eqref{LPKbar}. Assume we
know \eqref{LPK}$_{kl}$ for a fixed pair of indices $(k,l)$. As for
any $\{\alpha_j\}$ the matrix $\mathcal{U^*E}^t\mathcal U$ is a
doubly-infinite CMV matrix, we know that
\begin{equation}\label{rem1}
\{(\mathcal{U^*E}^t\mathcal U)_{kl},
K_{n+1}(\mathcal{U^*E}^t\mathcal U)\}=i[\mathcal{U^*E}^t\mathcal U,
(\mathcal{U^*E}^t\mathcal U)^{n+1}_+]_{kl}.
\end{equation}
But
\begin{align*}
K_{n+1}(\mathcal{U^*E}^t\mathcal U)&=\frac{1}{(n+1)d}{\rm
Tr}\big((\mathcal{U}_{(d)}^*\mathcal{Q}_{(d)}^t\mathcal
U_{(d)})^{n+1}\big)\\
&=\frac{1}{(n+1)d}{\rm Tr}\big(
\mathcal{U}_{(d)}^*(\mathcal{Q}_{(d)}^t)^{n+1}\mathcal U_{(d)}
\big)\\
&=K_{n+1}(\E)
\end{align*}
and $\mathcal U$ is a constant matrix. Therefore,
\begin{equation}\label{rem2}
\begin{aligned}
\{(\mathcal{U^*E}^t\mathcal U)_{kl},
K_{n+1}(\mathcal{U^*E}^t\mathcal U)\}&=\big(\mathcal U^* \{\E^t,
K_{n+1}(\E)\} \mathcal
U\big)_{kl}\\
&=\{\E^t_{k-1,l-1},K_{n+1}(\E)\}\\
&={\{\E_{l-1,k-1},K_{n+1}(\E)\}}.
\end{aligned}
\end{equation}
On the other hand,
\begin{equation}\label{rem3}
\begin{aligned}
i[\mathcal{U^*E}^t\mathcal U, (\mathcal{U^*E}^t\mathcal U)^n_+]_{kl}
&=i\big(\mathcal U^*[\E^t,(\E^t)^{n+1}_+]\mathcal
U\big)_{kl}=i[\E^t,(\E^t)^{n+1}_+]_{k-1,l-1}\\
&=i[\E^t,(\E^{n+1}_-)^t]_{k-1,l-1}=i[\E,\E^{n+1}_+]_{l-1,k-1}.
\end{aligned}
\end{equation}
Plugging in \eqref{rem2} and \eqref{rem3} into \eqref{rem1}, one
gets relation \eqref{LPK}$_{l-1,k-1}$:
$$
\{\E_{l-1,k-1}, K_{n+1}\}=i[\E, \E^{n+1}_+]_{l-1,k-1}.
$$
If instead of considering $\mathcal{U^*E}^t\mathcal U$ we consider
$\mathcal{UE}^t\mathcal U^*$, we obtain that \eqref{LPK}$_{kl}$
implies \eqref{LPK}$_{l+1,k+1}$. In particular, this means that:
\begin{itemize}
    \item \eqref{LPK}$_{kk}\quad\Leftrightarrow\quad$ \eqref{LPK}$_{k+1,k+1}$
    \item \eqref{LPK}$_{k,k-1}\quad\Leftrightarrow\quad$ \eqref{LPK}$_{k,k+1}$
    \item \eqref{LPK}$_{k+1,k-1}\quad\Leftrightarrow\quad$ \eqref{LPK}$_{k,k+2}$
    \item \eqref{LPK}$_{k+1,k}\quad\Leftrightarrow\quad$ \eqref{LPK}$_{k+1,k+2}$.
\end{itemize}
So the proof of relation \eqref{LPK} is complete once we prove it
for the indices $(k,k)$, $(k,k-1)$, $(k+1,k-1)$, and $(k+1,k)$ with
$k$ even.

We note here that we can apply the same reasoning as above to $\E^*$
instead of $\E^t$, but we do not obtain anything new.

Finally, these relations are proved using Lemma~\ref{PartialK}. We
give the computational details in Appendix~\ref{Comp}.

The second part of the proof deals with relation \eqref{LPKbar} for
$n+1$, $n\geq0$:
$$
\{\E,\bar{K}_{n+1}\}=[\E,i(\E^{n+1}_+)^*].
$$
We shall proceed in very much the same way as with \eqref{LPK},
while incorporating the necessary computational adjustments.

Let us first note that
$$
(\E^{n+1}_+)^*=((\E^*)^{n+1})_-.
$$
So Lemma~\ref{Shape} and the subsequent remarks apply here too and
we can conclude that $[\E,i(\E^{n+1}_+)^*]$ has the same shape as
$\E$.

Turning our attention to the entries inside the shape of $\E$, let
us note that using exactly the same reasoning as for
equation~\eqref{LPK} shows that
\begin{itemize}
    \item \eqref{LPKbar}$_{kk}\quad\Leftrightarrow\quad$ \eqref{LPKbar}$_{k+1,k+1}$
    \item \eqref{LPKbar}$_{k,k-1}\quad\Leftrightarrow\quad$ \eqref{LPKbar}$_{k,k+1}$
    \item \eqref{LPKbar}$_{k+1,k-1}\quad\Leftrightarrow\quad$ \eqref{LPKbar}$_{k,k+2}$
    \item \eqref{LPKbar}$_{k+1,k}\quad\Leftrightarrow\quad$ \eqref{LPKbar}$_{k+1,k+2}$.
\end{itemize}
So again we only have to check four relations; the only difference
is that, in this case, \eqref{LPKbar}$_{k+1,k+1}$,
\eqref{LPKbar}$_{k,k+1}$, \eqref{LPKbar}$_{k,k+2}$, and
\eqref{LPKbar}$_{k+1,k+2}$ turn out to be computationally easier to
verify. We do this in Appendix~\ref{Comp}.
\end{proof}

\section{The Finite Case}\label{Finite}

In this section we prove Lax pair representations for the finite
Ablowitz-Ladik system.

We are interested in studying the system
$$
-i\dot\alpha_j=\rho_j^2(\alpha_{j+1}+\alpha_{j-1})
$$
for $0\leq j\leq k-2$, with boundary conditions
$\alpha_{-1}=\alpha_{k-1}=-1$. The idea behind finding Lax pairs for
this system is to take one of the $\alpha$'s in the appropriate
periodic problem to the boundary, and identify all the objects
obtained in this way. As it turns out, they are all naturally
related to both the Ablowitz-Ladik system and orthogonal polynomials
on the circle, and can be defined independently of the periodic
setting.

Let us elaborate. As presented in Appendix~\ref{OPUC}, if we start
with a finitely supported measure $\mu$ on $S^1$, the associated
Verblunsky coefficients are
$\alpha_0,\ldots,\alpha_{k-2}\in\mathbb{D}$, $\alpha_{k-1}\in S^1$.
The CMV matrix is in this case a unitary $k\times k$ matrix
$$
\C_f=\LL_f\M_f
$$
with
$$
\LL_f=
\left(%
\begin{array}{ccccc}
  \bar\alpha_0 & \rho_0 &  &  &  \\
  \rho_0 & -\alpha_0 &  &  &  \\
   &  & \ddots &  &  \\
   &  &  & \bar\alpha_{k-2} & \rho_{k-2} \\
   &  &  & \rho_{k-2} & -\alpha_{k-2} \\
\end{array}%
\right)
$$
and
$$
\M_f=
\left(%
\begin{array}{ccccc}
  1 &  &  &  &  \\
   & \bar\alpha_1 & \rho_1 &  &  \\
   & \rho_1 & -\alpha_1 &  &  \\
   &  &  & \ddots &  \\
   &  &  &  & \bar\alpha_{k-1} \\
\end{array}%
\right).
$$
If, in addition, we restrict our attention to the case when
$\alpha_{k-1}=-1$, then we obtain the following connection between
the finite and the periodic cases:

\begin{lemma}
Let $k$ be even and $\C_f$ as above with $\alpha_{k-1}=-1$. Define a
doubly-infinite set of Verblunsky coefficients by periodicity:
$\alpha_{nk+j}=\alpha_j$ for all $n\in\mathbb{Z}$ and $0\leq j\leq
k-1$. Then the extended CMV matrix $\E$ associated to these
$\alpha$'s has the direct sum decomposition
\begin{equation}\label{decomp}
\E=\bigoplus_{r\in\mathbb{Z}}S^r(\C_f),
\end{equation}
where $S:l^{\infty}(\mathbb{Z})\rightarrow l^{\infty}(\mathbb{Z})$
is the right $k$-shift.

In particular, the following also hold:
\begin{equation}\label{decompd}
\Q_{(d)}=\bigoplus_{r=0}^{d-1} S^r(\C_f)
\end{equation}
and
\begin{equation}\label{Kfinite}
K_n(\E)=\frac{1}{n}{\rm{Tr}}(\C_f^n)
\end{equation}
for all $d\geq1$ and $n\geq1$.
\end{lemma}

\begin{proof}

Relation \eqref{decomp} follows immediately if we observe that
$\rho_{rk-1}=0$ for all $r\in\mathbb{Z}$, and this implies that (see
equation~\eqref{LMtilda})
$$
\tilde\M=\bigoplus_{r\in\mathbb{Z}}S^r(\M_f).
$$

By periodicity, we always have
$$
\tilde\LL=\bigoplus_{r\in\mathbb{Z}}S^r(\LL_f).
$$
So \eqref{decomp} follows from the definition of $\C_f=\LL_f\M_f$.
Likewise, \eqref{decompd} is just the restriction of \eqref{decomp}
to $X_{(d)}$. So then
$$
\Q^n_{(d)}=\bigoplus_{r=0}^{d-1} S^r(\C_f^n)
$$
and by taking the trace we get \eqref{Kfinite}.
\end{proof}

Note also that the Poisson bracket \eqref{DefPB} separates the
$\alpha$'s, and hence it naturally restricts to the space of
$(\alpha_0,\ldots,\alpha_{k-2},\alpha_{k-1}=-1)\in\mathbb{D}^{k-1}$.
If two functions $f$ and $g$ depend only on
$\alpha_0,\ldots,\alpha_{k-2}$, then
\begin{align*}
\{f,g\} &=\frac12 \sum_{j=0}^{k-2} \rho_j^2 \left[\frac{\partial
f}{\partial u_j}\frac{\partial g}{\partial v_j}-
 \frac{\partial f}{\partial v_j}\frac{\partial g}{\partial u_j}\right]\\
&=i\sum_{j=0}^{k-2} \rho_j^2 \left[\frac{\partial f}{\partial
\bar\alpha_j}\frac{\partial g}{\partial \alpha_j}-
 \frac{\partial f}{\partial \alpha_j}\frac{\partial g}{\partial
 \bar\alpha_j}\right],
\end{align*}
where, as before, $\alpha_j=u_j+i v_j$ for all $0\leq j\leq k-2$.

So the next theorem is an immediate consequence of Theorem~\ref{LP}:

\begin{theorem}\label{LPfinite}
Let
$$
K_n^f=K_n(\C_f)=\frac{1}{n}{\rm{Tr}}(\C_f^n)
$$
for all $n\geq1$. Then the Lax pairs associated to these
Hamiltonians are given by
\begin{equation}\label{LPfiniteK}
\{\C_f,K_n^f\}=[\C_f,i(\C_f^n)_+]
\end{equation}
and
\begin{equation}\label{LPfiniteKbar}
\{\C_f,\bar K_n^f\}=[\C_f,i((\C_f^n)_+)^*]
\end{equation}
for all $n\geq1$.

Or, in terms of real-valued flows, we have
\begin{equation}
\{\C_f,2\Re (K_n^f)\}=[\C_f,i(\C_f^n)_+ +i((\C_f^n)_+)^*]
\end{equation}
and
\begin{equation}
\{\C_f,2\Im (K_n^f)\}=[\C_f,(\C_f^n)_+ -((\C_f^n)_+)^*]
\end{equation}
for all $n\geq1$.
\end{theorem}

As in the periodic case, since $K_n^f$ is the trace of $\C_f^n$, we
obtain Poisson commutativity of the Hamiltonians:

\begin{coro} For all $m,n\geq1$, we have that
$$
\{K_n^f,\Re(K_m^f)\}=\{K_n^f,\Im(K_m^f)\}=0
$$
and
$$
\{K_n^f,K_m^f\}=\{K_n^f,\bar K_m^f\}=0.
$$
\end{coro}

\begin{remark}
Note that, since $\alpha_{k-1}\equiv-1$, we get $\rho_{k-1}\equiv0$,
and so $K_0=\prod_{j=0}^{k-1}\rho_j^2\equiv0$ on $\mathbb{D}^{k-1}$.
But if we define
$$
K_0^f=\prod_{j=0}^{k-2}\rho_j^2,
$$
then
$$
\{\alpha_m,K_0^f\}=iK_0^f\alpha_m,
$$
or
$$
\{\alpha_m,\log(K_0^f)\}=i\alpha_m.
$$
But, even though $K_0^f$ acts on the $\alpha$'s in the finite case
in the same way as $K_0$ does in the periodic case, there exists no
Lax pair representation for $K_0^f$ in terms of $\C_f$. The reason
is that
$$
\Tr\{\C_f,K_0^f\}=-iK_0^f(\bar\alpha_0-\alpha_{k-2}),
$$
which is not identically zero on $\mathbb{D}^{k-1}$, while the trace
of a commutator is always zero.
\end{remark}

\section{The Infinite Case}\label{Infinite}

Finally, we deal with the infinite defocusing Ablowitz-Ladik system.
By this, we mean that we consider the system whose first equation is
$$
i\dot\alpha_j=\rho_j^2 (\alpha_{j+1}+\alpha_{j-1})
$$
for all $j\geq0$, with the boundary condition $\alpha_{-1}=0$. The
idea behind constructing Lax pairs for this system is to use the
finite AL result. Since each entry in a fixed power of the CMV
matrix depends on only a bounded number of $\alpha$'s, extending the
finite Lax pairs to the infinite case only requires an appropriate
definition of the ``infinite" Hamiltonians $K_n^i$ for all $n\geq1$.

Let us explain these claims: Fix $n_0\geq1$ and $j_0,m_0\geq0$.
Consider the finite problem with $k$ very large ($k\geq
20(j_0+k_0+n_0)$ is sufficient, though a much more precise bound can
be found). In this case,
$$
\C^n_{j,m}=(\C_f)^n_{j,m}
$$
for all $0\leq j,m\leq j_0+4,m_0+4$ respectively, and $1\leq n\leq
n_0$. Say we can define a $K_n^i$ such that its dependence on the
first $k$ $\alpha$'s is the same as that of $K_n^f$. Then, for
$0\leq j,m\leq j_0+4,m_0+4$ respectively, we can replace ``finite"
by ``infinite" in \eqref{LPfiniteK}$_{j_0,m_0}$ and
\eqref{LPfiniteKbar}$_{j_0,m_0}$.

So the last element we need is $K_n^i$, the $n^{{\rm{th}}}$
Hamiltonian for the infinite problem. It is a function defined on
sequences $\{\alpha_j\}_{j\geq0}$ of numbers inside the unit disk,
having a certain decay. The condition that it must satisfy is that
$$
K_n^i(\{\alpha_0,\alpha_1\ldots,\alpha_{k-1}=-1,0,0,\ldots\})
=K_n^f(\{\alpha_0,\alpha_1,\ldots,\alpha_{k-1}=-1\}).
$$
Given that
$$
K_n^f(\C_f)=\frac{1}{n}{\rm{Tr}}(\C_f^n),
$$
a natural guess for $K_n^i$ would be
$$
K_n^i(\C)=\frac{1}{n}{\text{``Tr"}}(\C^n).
$$

But recall that the CMV matrix is unitary, so it is not trace class.
Nonetheless, given the special structure of $\C$, we can define our
Hamiltonian $K_n^i$ following this intuition as the sum of the
diagonal entries of $\C^n$. While this statement will be rigorously
proved in the following lemma, the reason why one can sum the series
of diagonal entries is that all of these entries have the same
structure for shifted $\alpha$'s: They are the sum of a bounded
number of ``monomials." By ``monomial" we mean a finite product of
$\alpha$'s and $\rho$'s. All the monomials that appear as terms in
the diagonal entries contain at least one $\alpha$ factor. Since all
the $\alpha$'s and $\rho$'s have absolute values less than 1, and if
we assume $l^1$-decay of the sequence of coefficients, the one
$\alpha$ factor in each monomial will ensure convergence of the
whole series.

The next Lemma and its proof explore in more detail the structure of
the entries of powers of the CMV matrix and its consequences for the
definition of Hamiltonians in the infinite case.

\begin{lemma}\label{L4.1}
Let $\{\alpha_j\}_{j\geq0}\in l^1(\mathbb{N})$ be a sequence of
coefficients with $\alpha_j\in\mathbb{D}$ for all $j\geq0$. Let $\C$
be the CMV matrix associated to these coefficients. Then the series
\begin{equation}\label{KinfSeries}
\sum_{k\geq0}\C^n_{k,k}
\end{equation}
converges absolutely for any $n\geq1$.

Moreover, for any $k\geq0$, we have that $\C^n_{k,k}$ depends only
on $\alpha_{k-(2n-1)}$, $\ldots,\alpha_{k+2n-1}$, where all the
$\alpha$'s with negative indices are assumed to be identically zero.
\end{lemma}

\begin{proof}
We prove these statements by making two important observations.

The first refers to the general, doubly-infinite case. Let
$\{\alpha_j\}_{j\in\mathbb{Z}}$ be a sequence of complex numbers in
$\mathbb{D}$, and $\E$ the associated extended CMV matrix. Notice
that the structure of $\E$ is such that there exist functions
$f_{1,d_1}^e$ and $f_{1,d_1}^o$ defined on $\mathbb{D}^3$ with
$$
\E_{j,k}=f_{1,d_1}^e(\alpha_{j-1},\alpha_j,\alpha_{j+1})
$$
for all $j$ even and $j-k=d_1$, and
$$
\E_{j+1,k}=f_{1,d_1}^o(\alpha_{j-1},\alpha_j,\alpha_{j+1})
$$
for all $j$ even and $(j+1)-k=d_1$. Here $e$ and $o$ are used to
denote ``even" or ``odd" respectively, and $-2\leq d_1\leq1$.

Using this simple remark, one can prove by induction that, for all
$n\geq1$, there exist functions
$$
f_{n,d_n}^{e},f_{n,d_n}^o : \mathbb{D}^{4n-1}\rightarrow \mathbb{C}
$$
with $-2n \leq d_n\leq 2n-1$ such that
$$
\E^n_{j,k}=f_{n,j-k}^e(\alpha_{j-(2n-1)},\ldots,\alpha_{j+(2n-1)})
$$
for $j$ even and $-2n\leq j-k\leq2n-1$,
$$
\E^n_{j+1,k}=f_{n,j-k+1}^o(\alpha_{j-(2n-1)},\ldots,\alpha_{j+(2n-1)})
$$
for $j$ even and $-2n\leq j+1-k\leq 2n-1$, and
$$
\E^n_{l,m}=0
$$
for all the other indices $(l,m)$.

Moreover, for $|d_n|\leq 2n-1$, each such function $f_{n,d_n}^{e/o}$
is a sum of at most $4^n$ monomials, that is, products of $\alpha$'s
and $\rho$'s, and each monomial contains at least one $\alpha$
factor. The only entries containing only $\rho$'s are the extreme
ones:
\begin{equation}\label{Eeven}
\begin{aligned}
\E^n_{j,j+2n}&=f_{n,-2n}^e(\alpha_{j-(2n-1)},\ldots,\alpha_{j+(2n-1)})\\
             &=\rho_{j}\rho_{j+1}\cdots\rho_{j+2n-1}
\end{aligned}
\end{equation}
and
\begin{equation}\label{Eodd}
\begin{aligned}
\E^n_{j+1,j-(2n-1)}&=f_{n,-2n}^o(\alpha_{j-(2n-1)},\cdots,\alpha_{j+(2n-1)})\\
                   &=\rho_{j-(2n-1)}\rho_{j-(2n-2)}\cdots\rho_j
\end{aligned}
\end{equation}
for all $j$ even.

Fix $n\geq1$. Each monomial in $f_{n,d_n}^{e/o}$ is bounded by the
absolute value of one of the $\alpha$'s involved, and there are
$4^n$ such monomial terms in each sum. Putting all of this together,
we get that, for all $j$ even, we have
$$
|\E^n_{j,j}|\, , \,|\E^n_{j+1,j+1}|
\leq4^n(|\alpha_{j-(2n-1)}|+\cdots+|\alpha_{j+2n-1}|).
$$

The second observation we need to make in order to conclude the
convergence of the series \eqref{KinfSeries} concerns what changes
in all of these formulae when we introduce a boundary condition
$\alpha_{-1}=-1$.

From the discussion above, we see that actually
$$
\C^n_{j,k}=\E^n_{j,k}
$$
for $j,k\geq 4n$, as these entries only depend on $\alpha$'s with
positive indices. (As we remarked earlier, these bounds are not
optimal, but they are certainly sufficient for our purposes.) Hence
we also get that
$$
|\C^n_{j,j}|\,,\,|\C^n_{j+1,j+1}|\leq4^n(|\alpha_{j-(2n-1)}|+\cdots+|\alpha_{j+2n-1}|)
$$
for $j\geq 4n$ even. So, since the sequence of $\alpha$'s is in
$l^1$, we get that, for any $n\geq1$, the series \eqref{KinfSeries}
converges absolutely.
\end{proof}

We can now define our Hamiltonians as
\begin{equation}\label{KinfDef}
K_n^i=K_n^i(\C)=\sum_{k=0}^{\infty} \C^n_{k,k}.
\end{equation}
They are well-defined by the previous lemma, and, for any fixed
$j\geq0$, only a finite number of terms in the series depends on
$\alpha_j$. We can therefore state our main theorem in the infinite
case:

\begin{theorem}\label{LPinfinite}
Let $\{\alpha_j\}_{j\geq0}$ be an $l^1(\mathbb{N})$ sequence of
complex numbers inside the unit disk, $\C$ the associated CMV
matrix, and $K_n^i$ the function defined by \eqref{KinfDef}. Then
the Lax pairs associated to these Hamiltonians are given by
\begin{equation}\label{LPinfiniteK}
\{\C,K_n^i\}=[\C,i\C^n_+]
\end{equation}
and
\begin{equation}\label{LPinfiniteKbar}
\{\C,\bar K_n^i\}=[\C,i(\C^n_+)^*]
\end{equation}
for all $n\geq1$.

Or, in terms of real-valued flows, we have
\begin{equation}
\{\C,2\Re (K_n^i)\}=[\C,i\C^n_+ +i(\C^n_+)^*]
\end{equation}
and
\begin{equation}
\{\C,2\Im (K_n^i)\}=[\C,\C^n_+ -(\C^n_+)^*]
\end{equation}
for all $n\geq1$.
\end{theorem}

\begin{proof}
For each fixed $n$ and entry $(j,l)$, there exists a $k$ large
enough such that all the entries of $\C$ and $\C^n$ that appear in
\eqref{LPinfiniteK}$_{j,l}$ and \eqref{LPinfiniteKbar}$_{j,l}$ are
equal to the entries of $\C_f$ and $\C_f^n$, respectively, in the
corresponding finite Lax pairs.

Moreover, since $\C_{j,l}$ depends on two $\alpha$'s, and these
appear in only finitely many of the terms in $K_n^i$, the Poisson
brackets on the left-hand side are well defined finite sums and
equal the corresponding Poisson brackets in the finite case.

Therefore, the results of Theorem~\ref{LPinfinite} follow directly
from Theorem~\ref{LPfinite} and the observations in the proof of
Lemma~\ref{L4.1}.
\end{proof}

\begin{remark}
As in the finite case, we define
$$
K_0^i=\prod_{j=0}^{\infty}\rho_j^2.
$$
Recall that
$$
\rho_j^2=1-|\alpha_j|^2\leq2(1-|\alpha_j|),
$$
and $\{\alpha_j\}_{j\geq0}\in l^1(\mathbb{N})$. Therefore $K_0^i$ is
well-defined and positive; also the following Poisson bracket makes
sense
$$
\{\alpha_j,\log(K_0^i)\}=-2i\alpha_j.
$$
But, as in the finite case, we cannot hope to find a Lax pair
representation for the flow generated by $K_0^i$ in terms of $\C$.
The dependence of $\sum_{j\geq0}\{\C_{jj},K_0^i\}$ on $\bar\alpha_0$
is nontrivial, while $\sum_{j\geq0}[\C,\mathcal{A}]_{jj}$ is
identically zero for any infinite matrix $\mathcal{A}$ for which the
commutator makes sense.
\end{remark}

\noindent \textit{Acknowledgments:} The author wishes to thank her
advisor, Barry Simon, for his encouragement and advice, and for
access to preliminary drafts of his forthcoming two-volume treatise,
\cite{Simon1} and \cite{Simon2}. She also thanks Percy Deift for
suggesting this problem, and Rowan Killip, for his very helpful
remarks on preliminary versions of this paper.

\appendix

\section{Theorem~\ref{LP}: The Full Computations}\label{Comp}

We prove relation \eqref{LPK} for the necessary indices.

First let $k=l$ be even. Then
\begin{equation*}
\begin{aligned}
i\{\E_{kk},K_{n+1}\} &=\sum_j \rho_j^2
\Big[\frac{\partial(-\alpha_{k-1}\bar\alpha_k)}{\partial
\alpha_j}\frac{\partial K_{n+1}}{\partial
\bar\alpha_j}-\frac{\partial(-\alpha_{k-1}\bar\alpha_k)}{\partial
\bar\alpha_j}\frac{\partial K_{n+1}}{\partial \alpha_j}\Big]\\
&=-\rho_{k-1}^2\bar\alpha_k\frac{\partial K_{n+1}}{\partial
\bar\alpha_{k-1}}+\rho_k^2\alpha_{k-1}\frac{\partial
K_{n+1}}{\partial\alpha_k}\\
&=-\rho_{k-1}^2\bar\alpha_k\Big[\rho_{k-2}\E^n_{k-1,k-2}
-\alpha_{k-2}\E^n_{k-1,k-1}
-\frac{\alpha_{k-1}\rho_{-2}}{2\rho_{k-1}} \E^n_{k,k-2}\\
&-\frac{\alpha_{k-1}\bar\alpha_{k}}{2\rho_{k-1}}\E^n_{k-1,k}
-\frac{\alpha_{k-1}\rho_{k}}{2\rho_{k-1}}\E^n_{k-1,k+1}
+\frac{\alpha_{k-1}\alpha_{k-2}}{2\rho_{k-1}}\E^n_{k,k-1}\Big]\\
&+\rho_k^2\alpha_{k-1}\Big[-\frac{\bar\alpha_k\bar\alpha_{k+1}}{2\rho_k}
\E^n_{k+1,k} -\frac{\bar\alpha_k\rho_{k+1}}{2\rho_k}\E^n_{k+2,k}
-\frac{\bar\alpha_k\rho_{k-1}}{2\rho_k}\E^n_{k-1,k+1}\\
&+\frac{\bar\alpha_k\alpha_{k-1}}{2\rho_k}\E^n_{k,k+1}
-\bar\alpha_{k+1}\E^n_{k+1,k+1} -\rho_{k+1}\E^n_{k+2,k+1}\Big]\,.
\end{aligned}
\end{equation*}
On the other hand,
\begin{equation*}
\begin{aligned}
\left[\E, \E^{n+1}_+\right]_{k,k}
&=\E_{k,k-1}\E^{n+1}_{k-1,k}-\E^{n+1}_{k,k+1}\E_{k+1,k}\\
&=\rho_{k-1}\bar\alpha_k\E^{n+1}_{k-1,k}+
\alpha_{k-1}\rho_k\E^{n+1}_{k,k+1}\\
&\begin{aligned}=\rho_{k-1}\bar\alpha_k \Big[
&\rho_{k-2}\rho_{k-1}\E^n_{k-1,k-2}-\alpha_{k-2}\rho_{k-1}\E^n_{k-1,k-1}\\
&-\alpha_{k-1}\bar\alpha_k\E^n_{k-1,k}-\alpha_{k-1}\rho_k\E^n_{k-1,k+1}\Big]
\end{aligned}\\
&\begin{aligned}+\alpha_{k-1}\rho_k \Big[
&\rho_{k-1}\bar\alpha_k\E^n_{k-1,k+1}-\alpha_{k-1}\bar\alpha_k\E^n_{k,k+1}\\
&+\rho_k\bar\alpha_{k+1}\E^n_{k+1,k+1}+\rho_k\rho_{k+1}\E^n_{k+2,k+1}\Big]\,.
\end{aligned}
\end{aligned}
\end{equation*}
After a few simple manipulations, we find that
$i\{\E_{kk},K_{n+1}\}+\left[\E, \E^{n+1}_+\right]_{kk}$ equals
\begin{equation*}
\begin{aligned}
&\begin{aligned} \frac{\alpha_{k-1}\bar\alpha_k}{2}\Big[
&\big(\E^n_{k,k-2}\rho_{k-2}\rho_{k-1}-\E^n_{k,k-1}\alpha_{k-2}\rho_{k-1}-\E^n_{k,k+1}\alpha_{k-1}\bar\alpha_k\big)\\
&-\big(\rho_{k-1}\bar\alpha_k\E^n_{k-1,k}+\rho_k\bar\alpha_{k+1}\E^n_{k+1,k}+\rho_k\rho_{k+1}\E^n_{k+2,k}\big)\Big]
\end{aligned}\\
&=\frac{\alpha_{k-1}\bar\alpha_k}{2}\Big[\big(\E^n\E\big)_{kk}-\big(\E\E^n\big)_{kk}\Big]=0,
\end{aligned}
\end{equation*}
which concludes the proof of \eqref{LPK}$_{kk}$.

\bigskip

The second case we must consider is \eqref{LPK}$_{k,k-1}$ with $k$
even. Again, we look first at
\begin{equation*}
\begin{aligned}
i\{\E_{k,k-1}, K_{n+1}\} &=\sum_j \rho_j^2
\Big[\frac{\partial(\rho_{k-1}\bar\alpha_k)}{\partial
\alpha_j}\frac{\partial K_{n+1}}{\partial
\bar\alpha_j}-\frac{\partial(\rho_{k-1}\bar\alpha_k)}{\partial
\bar\alpha_j}\frac{\partial K_{n+1}}{\partial \alpha_j}\Big]\\
&=\rho_{k-1}^2\left[-\frac{\bar\alpha_{k-1}\bar\alpha_k}{2\rho_{k-1}}
\frac{\partial K_{n+1}}{\partial\bar\alpha_{k-1}}
+\frac{\alpha_{k-1}\bar\alpha_k}{2\rho_{k-1}} \frac{\partial
K_{n+1}}{\partial \alpha_{k-1}}\right]
-\rho_{k-1}\rho_k^2\frac{\partial K_{n+1}}{\partial \alpha_k}\\
&
\begin{aligned}
=\frac{\rho_{k-1}\bar\alpha_k}{2}\Big[
&\rho_{k-2}\bar\alpha_{k-1}\E^n_{k-1,k-2}
-\alpha_{k-2}\bar\alpha_{k-1}\E^n_{k-1,k-1}\\
&-\alpha_{k-1}\bar\alpha_k\E^n_{k,k}
-\alpha_{k-1}\rho_k\E^n_{k,k+1}\Big]
\end{aligned}\\
&\begin{aligned}-\rho_{k-1}\rho_k\Big[&\rho_k\rho_{k+1}\E^n_{k+2,k+1}
+\rho_k\bar\alpha_{k+1}\E^n_{k+1,k+1}
-\frac{\alpha_{k-1}\bar\alpha_k}{2}\E^n_{k,k+1}\\
&\frac{\rho_{k-1}\bar\alpha_k}{2}\E^n_{k-1,k+1}
+\frac{\bar\alpha_k\rho_{k+1}}{2}\E^n_{k+2,k}
+\frac{\bar\alpha_k\bar\alpha_{k+1}}{2}\E^n_{k+1,k}\Big].\end{aligned}
\end{aligned}
\end{equation*}
On the other hand,
\begin{equation*}
[\E,\E^{n+1}_+]_{k,k-1}= -\rho_{k-1}\rho_k\E^{n+1}_{k,k+1}
+\frac{\rho_{k-1}\bar\alpha_k}{2}(\E^{n+1}_{k-1,k-1}-\E^{n+1}_{kk}).
\end{equation*}
If we write $\E^{n+1}=\E\E^n$ and plug in the appropriate entries in
the expression above, we obtain
\begin{equation*}
\begin{aligned}
& i\{\E_{k,k-1}, K_{n+1}\}+[\E,\E^{n+1}_+]_{k,k-1}\\
&\begin{aligned}=-\frac{\rho_{k-1}\bar\alpha_k}{2}\Big[
&\alpha_{k-1}\rho_k\E^n_{k,k+1}+ \rho_k \rho_{k+1} \E^n_{k+2,k}+
\rho_k \bar\alpha_{k+1} \E^n_{k+1,k}\\
&+ \rho_{k-1} \bar\alpha_k \E^n_{k-1,k} - \rho_{k-2} \rho_{k-1}
\E^n_{k,k-2} - \rho_{k-1} \bar\alpha_k \E^n_{k,k-1}\Big]
\end{aligned}\\
&=-\frac{\rho_{k-1} \bar\alpha_k}{2}\Big[(\E \E^n)_{kk}-(\E^n
\E)_{kk}\Big]=0,
\end{aligned}
\end{equation*}
which proves \eqref{LPK}$_{k,k-1}$.

Having done these two cases in some detail, we will just present the
main steps in the computations for \eqref{LPK}$_{k+1,k-1}$ and
\eqref{LPK}$_{k+1,k}$. A useful observation is that, since both
sides of our identities are polynomials in the $\alpha$'s and
$\bar\alpha$'s, one can more easily identify terms by keeping track
of the powers of $\frac{1}{2}$ that occur.

The right-hand side of \eqref{LPK}$_{k+1,k-1}$ gives us
$$
[\E,\E^{n+1}_+]_{k+1,k-1}= \frac{\rho_{k-1}\rho_k}{2} \big(
\E^{n+1}_{k-1,k-1} -\E^{n+1}_{k+1,k+1} \big).
$$
So these are the terms we want to identify on the left-hand side:
\begin{equation*}
\begin{aligned}
i\{\E_{k+1,k-1},K_{n+1}\} &=\sum_j \rho_j^2
\Big[\frac{\partial(\rho_{k-1}\rho_k)}{\partial
\alpha_j}\frac{\partial K_{n+1}}{\partial
\bar\alpha_j}-\frac{\partial(\rho_{k-1}\rho_k)}{\partial
\bar\alpha_j}\frac{\partial K_{n+1}}{\partial \alpha_j}\Big]\\
&\begin{aligned} =\frac{\rho_{k-1}\rho_k}{2}\Big[ &-\E^{n}_{k-1,k-2}
\rho_{k-2}\bar\alpha_{k-1}-\E^n_{k-1,k-1}(-\alpha_{k-2}\bar\alpha_{k-1})\\
&+(-\alpha_{k-1}\rho_k) \E^n_{k,k+1} -\E^n_{k-1,k}
\rho_{k-1}\bar\alpha_k \\
&+ (-\alpha_k\bar\alpha_{k+1})\E^n_{k+1,k+1} + (-\alpha_k\rho_{k+1})
\E^n_{k+2,k+1} \Big]
\end{aligned}\\
&\begin{aligned}+\frac{|\alpha_{k-1}|^2}{4}\Big[
&\rho_{k-2}\rho_k\E^n_{k,k-2}+ \bar\alpha_k\rho_k \E^n_{k-1,k}\\
&+\rho_k^2 \E^n_{k-1,k+1} - \alpha_{k-2}\rho_k \E^n_{k,k-1}\\
&-\rho_{k-2}\rho_k \E^n_{k,k-2} + \alpha_{k-2}\rho_k
\E^n_{k,k-1}\\
&-\bar\alpha_k\rho_k \E^n_{k-1,k} - \rho_k^2 \E^n_{k-1,k+1}\Big]
\end{aligned}\\
&\begin{aligned}+\frac{|\alpha_{k}|^2}{4}\Big[
&\rho_{k-1}\rho_{k+1}\E^n_{k+2,k}+ \rho_{k-1}
\bar\alpha_{k+1} \E^n_{k-1,k}\\
&+\rho_{k-1}^2 \E^n_{k-1,k+1} - \alpha_{k-1}\rho_{k-1}
\E^n_{k,k+1}\\
&-\rho_{k-1}\rho_{k+1} \E^n_{k+2,k} - \rho_{k-1}
\bar\alpha_{k+1} \E^n_{k+1,k}\\
&-\alpha_{k-1}\rho_{k-1} \E^n_{k,k+1} - \rho_{k-1}^2
\E^n_{k-1,k+1}\Big]
\end{aligned}\\
&\begin{aligned}=\frac{\rho_{k-1}\rho_k}{2}\Big[ &-\big( \E^n\E
\big)_{k-1,k-1} + \E^n_{k-1,k+1}\E_{k+1,k-1}\\
& + \big(\E\E^n\big)_{k+1,k+1} - \E_{k+1,k-1} \E^n_{k-1,k+1}\Big]
\end{aligned}\\
&=-[\E,\E^{n+1}_+]_{k+1,k-1}\,.
\end{aligned}
\end{equation*}

Finally, we deal with \eqref{LPK}$_{k+1,k}$. As before, we notice
that
\begin{equation*}
[\E, \E^{n+1}_+]_{k+1,k}= \rho_{k-1}\rho_k \E^{n+1}_{k-1,k} -
\frac{\alpha_{k-1}\rho_k}{2} \big( \E^{n+1}_{kk} -
\E^{n+1}_{k+1,k+1}\big).
\end{equation*}
The other side of the identity can be transformed as follows:
\begin{equation*}
\begin{aligned}
i\{\E_{k+1,k}, K_{n+1}\}
&=-\sum_j\rho_j^2\Big[\frac{\partial(\rho_{k}\alpha_{k-1})}{\partial
\alpha_j}\frac{\partial K_{n+1}}{\partial
\bar\alpha_j}-\frac{\partial(\rho_{k}\alpha_{k-1})}{\partial
\bar\alpha_j}\frac{\partial K_{n+1}}{\partial \alpha_j}\Big]\\
&\begin{aligned}= &-\rho_{k-1}\rho_k \Big[\E^n_{k-1,k-2}
\rho_{k-2}\rho_{k-1} + \E^n_{k-1,k-1}
(-\alpha_{k-2}\rho_{k-1})\Big]\\
&\begin{aligned}+\frac{\alpha_{k-1}\rho_k}{2}\Big[
&\rho_{k-2}\rho_{k-1} \E^n_{k,k-2} + \rho_{k-1}\bar\alpha_k
\E^n_{k-1,k}\\
&+\rho_{k-1}\rho_k \E^n_{k-1,k+1} - \alpha_{k-2}\rho_{k-1}
\E^N_{k,k-1}\\
&+\rho_{k-1}\bar\alpha_k \E^n_{k-1,k}
-\alpha_{k-1}\bar\alpha_k \E^n_{k,k}\\
&+\alpha_k\bar\alpha_{k+1} \E^n_{k+1,k+1} +\alpha_k\rho_{k+1}
\E^n_{k+2,k+1}\Big]
\end{aligned}\\
&\begin{aligned}+\frac{\alpha_{k-1}|\alpha_k|^2}{4}\Big[
&-\bar\alpha_{k+1} \E^n_{k+1,k} -\rho_{k+1}\E^n_{k+2,k} -\rho_{k-1} \E^n_{k-1,k+1}\\
&+\alpha_{k-1} \E^n_{k,k+1} +\bar\alpha_{k+1} \E^n_{k+1,k} +\rho_{k+1}\E^n_{k+2,k}\\
&+\rho_{k-1} \E^n_{k-1,k+1} -\alpha_{k-1} \E^n_{k,k+1}\Big]
\end{aligned}\\
\end{aligned}\\
&\begin{aligned}= &-\rho_{k-1}\rho_k \Big[ \E^n_{k-1,k-2}
\E_{k-2,k} +\E^n_{k-1,k-1} \E_{k-1,k}\Big]\\
&\begin{aligned}+\frac{\alpha_{k-1}\rho_k}{2}\Big[ &\E^n_{k,k-2}
\E_{k-2,k} +2\rho_{k-1}\bar\alpha_k \E^n_{k-1,k}\\
&-\E_{k+1,k-1} \E^n_{k-1,k+1} +2\rho_{k-1}\rho_k
\E^n_{k-1,k+1}\\
&+\E^n_{k,k-1} \E_{k-1,k} +\E^n_{k,k}
\E_{k,k}\\
&-\E_{k+1,k+1} \E^n_{k+1,k+1} -\E_{k+1,k+2} \E^n_{k+2,k+1} \Big]
\end{aligned}\\
\end{aligned}\\
&=-\rho_{k-1}\rho_k (\E^n \E)_{k-1,k} +\frac{\alpha_{k-1}\rho_k}{2}
\Big[ (\E^n \E)_{k,k} - (\E \E^n)_{k+1,k+1}\Big]\\
&=-[\E, \E^{n+1}_+]_{k+1,k}\,.
\end{aligned}
\end{equation*}
This concludes the proof of \eqref{LPK}$_{k+1,k}$, and hence of
relation \eqref{LPK}.

The second part of the proof deals with relation \eqref{LPKbar}:
$$
\{\E,\bar{K}_{n+1}\}=[\E,i(\E^{n+1}_+)^*].
$$
We shall proceed in very much the same way as with \eqref{LPK},
while incorporating the necessary computational adjustments.

Again, we only have to check four relations; the only difference is
that in this case \eqref{LPKbar}$_{k+1,k+1}$,
\eqref{LPKbar}$_{k,k+1}$, \eqref{LPKbar}$_{k,k+2}$, and
\eqref{LPKbar}$_{k+1,k+2}$ turn out to be computationally easier to
verify.

As before, we start with the diagonal entry,
\eqref{LPKbar}$_{k+1,k+1}$, that we shall prove in some detail.

We start by analyzing the left-hand side and observing that
\begin{equation*}
\begin{aligned}
i\{\E_{k+1,k+1},\bar K_{n+1}\}&=\sum_j \rho_j^2
\Big[\frac{\partial(-\alpha_k\bar\alpha_{k+1})}{\partial\alpha_j}\frac{\partial\bar
K_{n+1}}{\partial\bar\alpha_j}
-\frac{\partial(-\alpha_k\bar\alpha_{k+1})}{\partial\bar\alpha_j}\frac{\partial\bar
K_{n+1}}{\partial\alpha_j}\Big]\\
&=-\rho_k^2\bar\alpha_{k+1}\frac{\partial\bar K_{n+1}}{\partial
\bar\alpha_k} +\rho_{k+1}^2\alpha_k\frac{\partial\bar
K_{n+1}}{\partial\alpha_{k+1}}.
\end{aligned}
\end{equation*}
So by taking the complex conjugate in this relation, we get that
\begin{equation*}
\begin{aligned}
\overline{i\{\E_{k+1,k+1},\bar K_{n+1}\}}
&=-\rho_k^2\alpha_{k+1}\frac{\partial K_{n+1}}{\partial\alpha_k}
+\rho_{k+1}^2\bar\alpha_k\frac{\partial
K_{n+1}}{\partial\bar\alpha_{k+1}}\\
&\begin{aligned}=\rho_k\alpha_{k+1}\Big[
 &\rho_k\bar\alpha_{k+1}\E^n_{k+1,k+1}+\rho_k\rho_{k+1}\E_{k+2,k+1}^n\\
 &+\frac{\bar\alpha_k\bar\alpha_{k+1}}{2}\E^n_{k+1,k}+\frac{\bar\alpha_k\rho_{k+1}}{2}\E^n_{k+2,k}\\
 &+\frac{\bar\alpha_k\rho_{k-1}}{2}\E^n_{k-1,k+1}-\frac{\alpha_{k-1}\bar\alpha_k}{2}\E^n_{k,k+1}\Big]
 \end{aligned}\\
&\begin{aligned}+\bar\alpha_k\rho_{k+1}\Big[
 &\rho_k\rho_{k+1}\E^n_{k+1,k}-\alpha_k\rho_{k+1}\E^n_{k+1,k+1}\\
 &-\frac{\rho_k\alpha_{k+1}}{2}\E^n_{k+2,k}-\frac{\alpha_{k+1}\bar\alpha_{k+2}}{2}\E^n_{k+1,k+2}\\
 &-\frac{\alpha_{k+1}\rho_{k+2}}{2}\E^n_{k+1,k+3}+\frac{\alpha_k\alpha_{k+1}}{2}\E^n_{k+2,k+1}\Big].
 \end{aligned}
\end{aligned}
\end{equation*}

On the other hand, we have that
\begin{equation*}
\begin{aligned}
\overline{[\E,(\E^{n+1}_+)^*]_{k+1,k+1}} &=\sum_{k-1\leq j\leq
k+2}\overline{\E_{k+1,j}}(\E^{n+1}_+)_{k+1,j} -\sum_{k\leq j\leq
k+3} (\E^{n+1}_+)_{j,k+1}\overline{\E_{j,k+1}}\\
&=\overline{\E_{k+1,k+2}}\E^{n+1}_{k+1,k+2}-\E^{n+1}_{k,k+1}\overline{\E_{k,k+1}}\\
&=-\bar\alpha_k\rho_{k+1}\E^{n+1}_{k+1,k+2}-\rho_k\alpha_{k+1}\E^{n+1}_{k,k+1}.
\end{aligned}
\end{equation*}

Notice that
\begin{equation*}
\begin{aligned}
\overline{i\{\E_{k+1,k+1},\bar K_{n+1}\}}
&=-\rho_k^2\alpha_{k+1}\frac{\partial K_{n+1}}{\partial\alpha_k}
+\rho_{k+1}^2\bar\alpha_k\frac{\partial
K_{n+1}}{\partial\bar\alpha_{k+1}}\\
&\begin{aligned}=\rho_k\alpha_{k+1}\Big[
 &(\E\cdot\E^n)_{k,k+1}-\rho_{k-1}\bar\alpha_k\E^n_{k-1,k+1}+\alpha_{k-1}\bar\alpha_k\E^n_{k,k+1}\\
 &+\frac{\bar\alpha_{k}\bar\alpha_{k+1}}{2}\E^n_{k+1,k}+\frac{\bar\alpha_k\rho_{k+1}}{2}\E^n_{k+2,k}\\
 &+\frac{\bar\alpha_k\rho_{k-1}}{2}\E^n_{k-1,k+1}-\frac{\alpha_{k-1}\bar\alpha_k}{2}\E^n_{k,k+1}\Big]
 \end{aligned}\\
&\begin{aligned}+\bar\alpha_k\rho_{k+1}\Big[
 &(\E^n\cdot\E)_{k+1,k+2}+\alpha_{k+1}\bar\alpha_{k+2}\E^n_{k+1,k+2}+\alpha_{k+1}\rho_{k+2}\E^n_{k+1,k+3}\\
 &-\frac{\rho_k\alpha_{k+1}}{2}\E^n_{k+2,k}-\frac{\alpha_{k+1}\bar\alpha_{k+2}}{2}\E^n_{k+1,k+2}\\
 &-\frac{\alpha_{k+1}\rho_{k+2}}{2}\E^n_{k+1,k+3}+\frac{\alpha_k\alpha_{k+1}}{2}\E^n_{k+2,k+1}\Big].
 \end{aligned}
\end{aligned}
\end{equation*}

Therefore, we get that $\overline{i\{\E_{k+1,k+1},\bar K_{n+1}\}
+[\E,(\E^{n+1}_+)^*]_{k+1,k+1}}$ equals
\begin{equation*}
\begin{aligned}
&\begin{aligned} \frac{\bar\alpha_k\alpha_{k+1}}{2}\Big[
&-\rho_{k-1}\rho_k\E^n_{k-1,k+1}+\alpha_{k-1}\rho_k\E^n_{k,k+1}+\alpha_k\rho_{k+1}\E^n_{k+2,k+1}\\
&+\rho_k\bar\alpha_{k+1}\E^n_{k+1,k}+\rho_{k+1}\bar\alpha_{k+2}\E^n_{k+1,k+2}+\rho_{k+1}\rho_{k+2}\E^n_{k+1,k+3}\Big]
\end{aligned}\\
&=\frac{\bar\alpha_k\alpha_{k+1}}{2}\Big[-(\E\cdot\E^n)_{k+1,k+1}+(\E^n\cdot\E)_{k+1,k+1}\Big]=0,\\
\end{aligned}
\end{equation*}
which ends the proof of \eqref{LPKbar}$_{k+1,k+1}$.

We now turn to \eqref{LPKbar}$_{k,k+1}$:
\begin{equation*}
\overline{[\E,(\E^{n+1}_+)^*]}_{k,k+1}=
\rho_k\rho_{k+1}\E^{n+1}_{k+1,k+2}
+\frac{\rho_k\alpha_{k+1}}{2}\Big(\E^{n+1}_{k+1,k+1}-\E^{n+1}_{k,k}\Big).
\end{equation*}
The left-hand side becomes
\begin{equation*}
\begin{aligned}
\overline{i\{\E_{k,k+1},\bar K_{n+1}\}}&=\overline{i\{\rho_k\bar\alpha_{k+1},\bar K_{n+1}\}}\\
&\begin{aligned}=&-\rho_k\rho_{k+1}\Big(\E^n_{k+1,k}\E_{k,k+2} +\E^n_{k+1,k+1}\E_{k+1,k+2}\Big)\\
 &\begin{aligned}-\frac{\rho_k\alpha_{k+1}}{2}\Big(
  &-\E_{k,k+2}\E^n_{k+2,k} +\E^n_{k+1,k+3}\E_{k+3,k+1}\\
  &-2\E^n_{k+1,k+3}\rho_{k+1}\rho_{k+2}
  +\E^n_{k+1,k+2}\E_{k+2,k+1}\\
  &-2\E^n_{k+1,k+2}\rho_{k+1}\bar\alpha_{k+2} -\E_{k,k-1}\E^n_{k-1,k}\\
  &-\E_{k,k}\E^n_{k,k}+\E^n_{k+1,k+1}\E_{k+1,k+1}\Big)
  \end{aligned}
 \end{aligned}\\
&=-\rho_k\rho_{k+1}(\E^n\E)_{k,k+1}
-\frac{\rho_k\alpha_{k+1}}{2}\Big((\E\E^n)_{k,k}+(\E^n\E)_{k+1,k+1}\Big).
\end{aligned}
\end{equation*}
So we find what we wanted:
$$
\{\E_{k,k+1},\bar K_{n+1}\}=i[\E,(\E^{n+1}_+)^*]_{k,k+1}.
$$

The next entry that we analyze is \eqref{LPKbar}$_{k,k+2}$.
Considering the right-hand side first, we get
\begin{equation*}
\begin{aligned}
\overline{[\E,(\E^{n+1}_+)^*]_{k,k+2}}
&=\sum_{j}\overline{\E_{k,j}}\cdot(\E^{n+1}_+)_{k+2,j}-\sum_j(\E^{n+1}_+)_{j,k}\cdot\overline{\E_{j,k+2}}\\
&=\frac{\rho_k\rho_{k+1}}{2}\big(\E^{n+1}_{k+2,k+2}-\E^{n+1}_{k,k}\big).
\end{aligned}
\end{equation*}

If we look at the left-hand side now, we get
\begin{equation*}
\begin{aligned}
\overline{i\{\E_{k,k+2},\bar K_{n+1}\}}
&=\overline{i\{\rho_k\rho_{k+1},\bar K_{n+1}\}}\\
&\begin{aligned}
 &=\rho_k^2\Big[-\frac{\alpha_k\rho_{k+1}}{2\rho_k}\cdot\frac{\partial K_{n+1}}{\partial\alpha_{k}}
  +\frac{\bar\alpha_k\rho_{k+1}}{2\rho_k}\cdot\frac{\partial
  K_{n+1}}{\partial\bar\alpha_k}\Big]\\
 &+\rho_{k+1}^2\Big[-\frac{\alpha_{k+1}\rho_k}{2\rho_{k+1}}\cdot\frac{\partial K_{n+1}}{\partial\alpha_{k+1}}
  +\frac{\bar\alpha_{k+1}\rho_k}{2\rho_{k+1}}\cdot\frac{\partial K_{n+1}}{\partial\bar\alpha_{k+1}}\Big]
 \end{aligned}\\
&\begin{aligned}
 &\begin{aligned}=\frac{\rho_k\rho_{k+1}}{2}\Big[
  &\alpha_k\rho_{k+1}\E^n_{k+2,k+1}+\bar\alpha_k\rho_{k-1}\E^n_{k-1,k}\\
  &-\alpha_{k-1}\bar\alpha_k\E^n_{k,k}+\alpha_{k+1}\bar\alpha_{k+2}\E^n_{k+2,k+2}\\
  &+\alpha_{k+1}\rho_{k+2}\E^n_{k+2,k+3}+\bar\alpha_{k+1}\rho_k\E^n_{k+1,k}\Big]
  \end{aligned}\\
 &\begin{aligned}+\frac{|\alpha_k|^2\rho_{k+1}}{4}\Big[
  &\bar\alpha_{k+1}\E^n_{k+1,k}+\rho_{k+1}\E^n_{k+2,k}\\
  &+\rho_{k-1}\E^n_{k-1,k+1}-\alpha_{k-1}\E^n_{k,k+1}\\
  &-\bar\alpha_{k+1}\E^n_{k+1,k}-\rho_{k+1}\E^n_{k+2,k}\\
  &-\rho_{k-1}\E^n_{k-1,k+1}+\alpha_{k-1}\E^n_{k,k+1}\Big]
  \end{aligned}\\
 &\begin{aligned}+\frac{|\alpha_{k+1}|^2\rho_k}{4}\Big[
  &\rho_k\E^n_{k+2,k}-\alpha_k\E^n_{k+2,k+1}\\
  &+\bar\alpha_{k+2}\E^n_{k+1,k+2}+\rho_{k+2}\E^n_{k+1,k+3}\\
  &-\rho_k\E^n_{k+2,k}+\alpha_k\E^n_{k+2,k+1}\\
  &-\bar\alpha_{k+2}\E^n_{k+1,k+2}-\rho_{k+2}\E^n_{k+1,k+3}\Big]
  \end{aligned}
 \end{aligned}\\
&\begin{aligned}=\frac{\rho_k\rho_{k+1}}{2}\Big[
 &(\E\E^n)_{k,k}-\E_{k,k+2}\E^n_{k+2,k}\\
 &-(\E^n\E)_{k+2,k+2}+\E^n_{k+2,k}\E_{k,k+2}\Big]
 \end{aligned}\\
&=-\overline{[\E,(\E^{n+1}_+)^*]_{k,k+2}}\,,
\end{aligned}
\end{equation*}
which immediately implies the equation \eqref{LPKbar}$_{k,k+2}$.

Finally, we turn to the last relation we have to prove, equation
\eqref{LPKbar}$_{k+1,k+2}$. As above, we start with the right-hand
side and observe that
\begin{equation*}
\overline{[\E,(\E^n_+)^*]}_{k+1,k+2}=\frac{\bar\alpha_k\rho_{k+1}}{2}\big(\E^{n+1}_{k+1,k+1}-\E^{n+1}_{k+2,k+2}\big)
-\rho_k\rho_{k+1}\E^{n+1}_{k,k+1}.
\end{equation*}

Considering the left-hand side now, we have
\begin{equation*}
\begin{aligned}
\overline{i\{\E_{k+1,k+2},\bar K_{n+1}\}}
&\begin{aligned}=&-\rho_k^2\rho_{k+1}\frac{\partial
K_{n+1}}{\partial\alpha_k}\\
 &-\bar\alpha_k\rho_{k+1}^2
 \Big[-\frac{\alpha_{k+1}}{2\rho_{k+1}}\frac{\partial
 K_{n+1}}{\partial\alpha_{k+1}}
 +\frac{\bar\alpha_{k+1}}{2\rho_{k+1}}\frac{\partial
 K_{n+1}}{\partial\bar\alpha_{k+1}}\Big]
 \end{aligned}\\
&=\rho_k\rho_{k+1}\Big[\rho_k\bar\alpha_{k+1}\E^n_{k+1,k+1}+\rho_k\rho_{k+1}\E^n_{k+2,k+1}\Big]\\
&\begin{aligned}-\frac{\bar\alpha_k\rho_{k+1}}{2}\Big[
 &-\rho_k\bar\alpha_{k+1}\E^n_{k+1,k}-\rho_k\rho_{k+1}\E^n_{k+2,k}\\
 &-\rho_k\rho_{k-1}\E^n_{k-1,k+1}+\alpha_{k-1}\rho_k\E^n_{k,k+1}\\
 &+\alpha_{k+1}\bar\alpha_{k+2}\E^n_{k+2,k+2}+\alpha_{k+1}\rho_{k+2}\E^n_{k+2,k+3}\\
 &+\bar\alpha_{k+1}\rho_k\E^n_{k+1,k}-\alpha_k\bar\alpha_{k+1}\E^n_{k+1,k+1}\Big]
 \end{aligned}\\
&\begin{aligned}-\frac{\bar\alpha_k|\alpha_{k+1}|^2}{4}\Big[
 &\rho_k\E^n_{k+2,k}-\alpha_k\E^n_{k+2,k+1}\\
 &+\bar\alpha_{k+2}\E^n_{k+1,k+2}+\rho_{k+2}\E^n_{k+1,k+3}\\
 &-\rho_k\E^n_{k+2,k}-\bar\alpha_{k+2}\E^n_{k+1,k+2}\\
 &-\rho_{k+2}\E^n_{k+1,k+3}+\alpha_k\E^n_{k+2,k+1}\Big]
 \end{aligned}\\
&=\rho_k\rho_{k+1}\Big[\rho_k\bar\alpha_{k+1}\E^n_{k+1,k+1}+\rho_k\rho_{k+1}\E^n_{k+2,k+1}\Big]\\
&\begin{aligned}-\frac{\bar\alpha_k\rho_{k+1}}{2}\Big[
 &-(\E^n\E)_{k+2,k+2}+(\E\E^n)_{k+1,k+1}\\
 &-2\rho_{k-1}\rho_k\E^n_{k-1,k+1}+2\alpha_{k-1}\rho_k\E^n_{k,k+1}\Big]
 \end{aligned}\\
&=\rho_k\rho_{k+1}\E^{n+1}_{k,k+1}-\frac{\bar\alpha_k\rho_{k+1}}{2}\big(\E^{n+1}_{k+1,k+1}-\E^{n+1}_{k+2,k+2}\big)\\
&=-\overline{[\E,(\E^n_+)^*]}_{k+1,k+2}\,,
\end{aligned}
\end{equation*}
which implies that
$$
i\{\E_{k+1,k+2},\bar K_{n+1}\}=-[\E,(\E^n_+)^*]_{k+1,k+2},
$$
and hence \eqref{LPKbar}$_{k+1,k+2}$ holds.

\section{Background: Orthogonal Polynomials on the Unit Circle}\label{OPUC}

In this Appendix we present some of the basic notions and results
related to the theory of orthogonal polynomials on the unit circle.
The reader interested in more details can check Szeg\H o's classical
book \cite{Szego}. In our presentation, we follow the upcoming
two-volume treatise by Simon~\cite{Simon1,Simon2}.

Let us first recall the definition of the Verblunsky coefficients.
Consider a probability measure $d\mu$ on $S^1$ which is supported at
infinitely many points. By applying the Gram-Schmidt procedure to
$1,z,z^2,\ldots$, one obtains the monic orthogonal polynomials
$\{\Phi_n(z)\}_{n\geq0}$ and the orthonormal polynomials
$$
\phi_n(z)=\frac{\Phi_n(z)}{\|\Phi_n\|_{L^2(d\mu)}}.
$$
These polynomials obey recurrence relations
\begin{align}
\Phi_{k+1}(z)   &= z\Phi_k(z)   - \bar{\alpha}_k \Phi_k^*(z),
\label{PhiRec}\\
\Phi_{k+1}^*(z) &=  \Phi_k^*(z) -    \alpha_k z  \Phi_k(z),
\label{Phi*Rec}
\end{align}
where the $\alpha_k$'s are the recurrence coefficients and
$\Phi_k^*$ denotes the reversed polynomial:
\begin{equation}\label{rev}
\Phi_k(z) = \sum_{l=0}^k c_l z^l \quad \Rightarrow \quad \Phi_k^*(z)
= \sum_{l=0}^k \bar{c}_{k-l} z^l.
\end{equation}
Equivalently, $\Phi_k^*(z) = z^k \overline{\Phi_k(\bar{z}^{-1})}$.
These recurrence equations imply
\begin{equation}\label{PhiNorm}
\bigl\|\Phi_k\bigr\|_{L^2(d\mu)} = \prod_{l=0}^{k-1}
\rho_l\quad\text{where}\quad
    \rho_l=\sqrt{1-|\alpha_l|^2},
\end{equation}
from which the recurrence relations for the orthonormal polynomials
are easily derived. The recurrence coefficients $\alpha_k$ are
called Verblunsky coefficients and lie in the (open) unit disk
$\mathbb{D}$.

The recursion relations for the orthonormal polynomials can be
summarized as
$$
\begin{bmatrix}
  \phi_n(z) \\
  \phi_n^*(z) \\
\end{bmatrix}
=A(\alpha_{n-1},z)
\begin{bmatrix}
  \phi_{n-1}(z) \\
  \phi_{n-1}^*(z) \\
\end{bmatrix},
$$
where
$$
A(\alpha_{k},z)=\frac{1}{\rho_k}
\begin{bmatrix}
z                &   -\bar{\alpha}_k\\
-\alpha_k z      &   1\\
\end{bmatrix}.
$$
We define the transfer matrix
\begin{equation}\label{transf}
T_n(z)=A(\alpha_{n-1},z)\ldots A(\alpha_0,z)
\end{equation}
for all $n\geq1$; hence
$$
\begin{bmatrix}
  \phi_n(z) \\
  \phi_n^*(z) \\
\end{bmatrix}
=T_n(z)
\begin{bmatrix}
  1 \\
  1 \\
\end{bmatrix}.
$$

Consider the operator $f(z)\mapsto zf(z)$ in $L^2(d\mu)$. We want to
represent this operator as a matrix. The most obvious choice of an
orthonormal set of vectors in $L^2(d\mu)$ are the orthonormal
polynomials, $\{\phi_n\}_{n\geq0}$. This leads to a matrix whose
entries can be expressed simply in terms of the $\alpha$'s. However,
this matrix is typically not sparse: All entries above and including
the sub-diagonal are non-zero; it is also unclear how to extend this
matrix to a doubly-infinite matrix, which will turn out to be very
important when we consider the case of periodic Verblunsky
coefficients. Moreover, $\{\phi_n\}_{n\geq0}$ is a basis in
$L^2(d\mu)$ if and only if $\{\alpha_j\}_{j\geq0}$ are not in
$l^2(\mathbb{N})$.

An alternate approach, due to Cantero, Moral, and Vel\'azquez
\cite{CMV}, consists of defining two bases in $L^2(d\mu)$. Applying
the Gram--Schmidt procedure to
$$
1,z,z^{-1},z^2,z^{-2},\ldots
$$
in $L^2(d\mu)$ produces the orthonormal basis
\begin{equation}
\chi_k(z) = \begin{cases}
    z^{-k/2}    \phi_k^*(z), & \text{$k$ even;} \\
    z^{(1-k)/2} \phi_k(z), & \text{$k$ odd,}\end{cases}
\end{equation}
where $k\geq0$.  As above, $\phi_k$ denotes the $k^{\rm{th}}$
orthonormal polynomial and $\phi_k^*$, its reversal
(cf.~\eqref{rev}). If we apply the procedure to
$1,z^{-1},z,z^{-2},z^2,\ldots$, instead, then we obtain a second
orthonormal basis:
\begin{equation}
x_k(z)=\overline{\chi_k(1/\bar z)}= \begin{cases}
    z^{-k/2}    \phi_k(z), & \text{$k$ even;} \\
    z^{(-1-k)/2} \phi_k^*(z), & \text{$k$ odd.}\end{cases}
\end{equation}

It is natural to compute the matrix representation of $f(z)\mapsto
zf(z)$ in $L^2(d\mu)$ with respect to these bases. The matrices with
entries
$$
\mathcal{L}_{i+1,j+1} = \langle \chi_i(z)| z x_j(z) \rangle
\quad\text{and}\quad \mathcal{M}_{i+1,j+1} = \langle   x_i(z) |
\chi_j(z)\rangle
$$
are block-diagonal; indeed,
\begin{equation}\label{LMdefn}
\mathcal{L}=\diag\bigl(\Theta_0   ,\Theta_2,\Theta_4,\ldots\bigr)
\quad\text{and}\quad
\mathcal{M}=\diag\bigl([1],\Theta_1,\Theta_3,\ldots\bigr),
\end{equation}
where
$$
\Theta_k = \begin{bmatrix} \bar\alpha_k & \rho_k \\ \rho_k &
-\alpha_k
\end{bmatrix}.
$$
The representation of $f(z)\mapsto zf(z)$ in the $\{\chi_j\}$ basis
is just
$$
\mathcal{C}=\mathcal{LM}=
\left(%
\begin{array}{cccccc}
  \bar\alpha_0 & \rho_0\bar\alpha_1    & \rho_0\rho_1    & 0 & 0 & \ldots \\
  \rho_0       & -\alpha_0\bar\alpha_1 & -\alpha_0\rho_1 & 0 & 0 & \ldots \\
  0 & \rho_1\bar\alpha_2 & -\alpha_1\bar\alpha_2 & \rho_2\bar\alpha_3 & \rho_2\rho_3 & \ldots \\
  0 & \rho_1\rho_2 & -\alpha_1\rho_2 & -\alpha_2\bar\alpha_3 & -\alpha_2\rho_3 & \ldots \\
  0 & 0 & 0 & \rho_3\bar\alpha_4 & -\alpha_3\bar\alpha_4 & \ldots \\
  \ldots & \ldots & \ldots & \ldots & \ldots & \ldots \\
\end{array}%
\right),
$$
which is called the CMV matrix, and in the $\{x_j\}$ basis, it is
$\tilde{\mathcal{C}}=\mathcal{ML}$. Let us note here that throughout
the paper we index rows and columns of matrices starting with 0: for
example $\LL_{jj}=\bar\alpha_j$ for all $j\geq0$. The (infinite) CMV
matrix $\C$ is the matrix that we use in Section~\ref{Finite} to
define Lax pairs for the flows generated by the Ablowitz-Ladik
Hamiltonians on the coefficients $\alpha_j$, $j\geq0$.

If the probability measure $\mu$ on the circle is supported at $k-1$
points, then we can define as above the orthogonal polynomials
$\{\Phi_n(z)\}_{0\leq n\leq k-2}$ and the corresponding orthonormal
polynomials $\{\phi_n(z)\}_{0\leq n\leq k-2}$. They still obey the
same recurrence relations, which allow us to identify the Verblunsky
coefficients $\alpha_0,\ldots,\alpha_{k-2}\in\mathbb{D}$ and
$\alpha_{k-1}\in S^1$. If, as in the infinite case, we represent the
operator of multiplication by $z$ in the basis considered by
Cantero, Moral, and Vel\'azquez we obtain a finite CMV matrix
$$
\C_f=\LL_f\M_f\,.
$$
Note that, since $|\alpha_{k-1}|=1$,
$$
\Theta_{k-1} = \begin{bmatrix} \bar\alpha_{k-1} & 0 \\ 0 &
-\alpha_{k-1}
\end{bmatrix}
$$
decomposes as the direct sum of two $1\times1$ matrices. Hence, if
we replace $\Theta_{k-1}$ by the $1\times 1$ matrix that is its top
left entry, $\bar\alpha_{k-1}$, and discard all $\Theta_m$ with
$m\geq k$, we find that $\LL_f$ and $\M_f$ are naturally $k\times k$
block-diagonal matrices. As in the infinite case, the finite CMV
matrix $\C_f$ allows us to recast the Ablowitz-Ladik hierarchy of
equations in Lax pair form.

Now we turn to the case of periodic Verblunsky coefficients. If the
$\alpha$'s are periodic with period $p$ even, that is, they obey
$\alpha_{j+p}=\alpha_j$ for all $j\geq0$, then we can define a
two-sided infinite sequence of coefficients by periodicity. The
extended CMV matrix is
\begin{equation}\label{DefE}
\mathcal{E}=\mathcal{\tilde{L}\tilde{M}},
\end{equation}
where
\begin{equation}\label{LMtilda}
\mathcal{\tilde L}=\bigoplus_{j\,\,\text{even}} \Theta_j \quad\quad
\text{and} \quad\quad \mathcal{\tilde M}=\bigoplus_{j\,\,\text{odd}}
\Theta_j,
\end{equation}
with $\Theta_j$ defined on $l^2(\mathbb{Z})$ by
$$
\Theta_j = \begin{bmatrix} \bar\alpha_j & \rho_j \\ \rho_j &
-\alpha_j
\end{bmatrix}
$$
on the span of $\delta_j$ and $\delta_{j+1}$, and identically 0
otherwise. The extended CMV matrix $\mathcal{E}$ will play an
important role in determining the Lax pairs associated to the
Hamiltonian flows of the periodic Ablowitz-Ladik system.


\end{document}